\documentclass{article}

\usepackage{amsmath}
\usepackage{graphicx}
\usepackage{amsfonts}
\usepackage{amssymb}

\def\binom#1#2{{#1 \choose #2}}
\newtheorem{theor}{Theorem}[section]
\newtheorem{cor}{Corrolary}[section]
\newtheorem{pr}{Proposition}[section]

\begin{document}

\title{Affine Hamiltonians in higher order geometry\thanks{%
The paper is published in Int J Theor Phys 46, 10 (2007), 2531--2549. The
original publication is available at www.springerlink.com}}
\author{Paul Popescu and Marcela Popescu}
\date{}
\maketitle

\begin{abstract}
Affine hamiltonians are defined in the paper and their study is based
especially on the fact that in the hyperregular case they are dual objects
of lagrangians defined on affine bundles, by mean of natural Legendre maps.
The variational problems for affine hamiltonians and lagrangians of order $%
k\geq2$ are studied, relating them to a Hamilton equation. An Ostrogradski
type theorem is proved: the Hamilton equation of an affine familtonian $h$
is equivalent with Euler-Lagrange equation of its dual lagrangian $L$.
Zermelo condition is also studied and some non-trivial examples are given. 
\newline

\end{abstract}

\thispagestyle{empty}



\section{Introduction}

Higher order lagrangians and hamiltonians are considered by Ostrogradski in
the study of equivalence problem of Euler-Lagrange and Hamilton equations of
a hyperregular lagrangian. A modern form of these ideas is considered in 
\cite[Chap.3, Sect.1.4]{AG1}, using the higher order tangent bundles $T^{k}M$
of a manifold $M$, where the dual hamiltonian defined by a strictly convex
lagrangian of higher order and the Hamilton equation are considered both on $%
T^{\ast }T^{k-1}M$. In \cite{Le}, one uses also the bundles of accelerations
in a systematic study of mechanical systems, using the Klein formalism. A
theory of higher order Hamilton spaces was recently studied in \cite{MiHS},
but the duality hamiltonian-lagrangian is not canonical and the action of
the lagrangian and its dual hamiltonian are not related by canonical
Legendre maps. An affine framework for lagrangians and hamiltomians can be
found in \cite{GGU, IMPS, U1} or \cite{PMP}. In order to try other methods
one can follow similar ideas used in \cite{CI}.

In this paper we consider a new definition, that of an affine hamiltonian of
higher order $k\geq2$ on a manifold $M$. An affine hamiltonian is an affine
section in an affine bundle with a one dimensional fiber and it is studied
using local real functions (i.e. local hamiltonians). These local
hamiltonians were considered by Arnold-Givental, in the case when the
lagrangian is convex (see Proposition \ref{prpoz}), but not stressing their
local form. An affine hamiltonian and a hamiltonian considered in \cite{MiHS}
(called in our paper a vectorial hamiltonian) are related by an affine
section.

We define the energy of an affine hamiltonian as a global hamiltonian (of
order one) on $T^{k-1}M$. The action of an affine hamiltonian is defined on
curves on $T^{\ast}M$ an it has as critical curves that given by the
Hamilton equation of the energy. Considering a hyperregular lagrangian of
order $k\geq2$, a canonical duality lagrangian-affine hamiltonian is
constructed by natural Legendre/Legendre$^{\ast}$ maps. The actions on
curves of the hamiltonian and its dual lagrangian are related by a
Ostrogradski type theorem: a hyperregular lagrangian $L$ and its dual affine
hamiltonian $h$ of order $k$ have the same energy, thus the same hamiltonian
vector field (Theorem \ref{the}) and the Hamilton equation oh $h$ and the
Euler-Lagrange equations of $L$ have the same solutions as curves on $M$
(Corollary\ref{corlagham}). The Ostrogradski theorem stated in \cite[Chap.3,
Sect.1.4]{AG1} for a strictly convex lagrangian is extended in this way to a
hyperregular lagrangian of order $k$. The action on curves is different from
that considered in \cite{MiHS}, but we investigate here some similar Zermelo
conditions; we find that there are no general Zermelo conditions for affine
hamiltonians (Proposition \ref{prZer}).

The consideration of an affine hamiltonian allows to perform a construction
of some Legendre and Legendre$^{\ast}$ maps associated with a Lagrangian and
an affine hamiltonian respectively, without using affine sections. The use
of an affine section in the construction of Legendre maps, as in \cite{MiHS,
Mi-ha-B}, makes the Legendre and Legendre$^{\ast}$ maps non-canonical
associated to lagrangians and hamiltonians, thus our construction improves
this aspect. The integral action of a (vectorial) hamiltonian of order $k$
on $M$, as defined and studied in \cite[Chapter 5]{Mi-ha-B}, seems to be not
related, in the hyperregular case, to any integral action of a lagrangian of
order $k$ on $M$. Notice also that the Hamilton equation obtained in \cite[%
Chapter 5]{Mi-ha-B} is completely different as that obtained bellow, and its
solutions as well.

In the first section we briefly discuss the Legendre maps between vectorial
and affine hamiltonians on one way and lagrangians on the other way, defined
on open sets of affine spaces.

A recursive definition of the $k$-tangent spaces (as affine bundles) is
performed in Section 2. We stress the role played by a vector pseudofield in
the construction of $T^{k}M$, since a similar ideea can be followed in order
to study other cases (for example to construct a time dependent $k$-tangent
space; notice that the case studied in this paper is time independent).

The variational problems for affine hamiltonians and lagrangians of order $%
k\geq2$ are studied in Section 3 and Section 5 respectively, relating them
to the Hamilton equation. The Zermelo condition is studied in Section 4.
Some non-trivial examples are considered in Section 5.

\section{Vectorial and affine Hamiltonians and \newline
Lagrangians on affine spaces}

Let $\mathcal{A}$ be a real affine space, modelled on a real vector space $V$%
. The \emph{vectorial dual} of $\mathcal{A}$ is $\mathcal{A}^{\dagger }=Aff(%
\mathcal{A},I\!\!R)$, where $Aff$ denotes affine morphisms. An \emph{affine
frame} on $\mathcal{A}$ is a couple $\mathcal{R}=(o,\mathcal{B})$, where $%
o\in\mathcal{A}$ is a point and $\mathcal{B}=\{\bar{e}_{i}\}_{i=\overline{1,n%
}}\subset V$ is a vector base. If $z\in\mathcal{A}$ is an arbitrary point,
then its \emph{affine coordinates} (or simply \emph{coordinates}) in this
frame, are the coordinates $(z^{i})_{i=\overline {1,n}}$ of the vector $%
\overline{oz}$, i.e. $\overline{oz}=z^{i}\bar{e}_{i}$.

Consider now an affine frame $\mathcal{R}=(o,\mathcal{B})$ of $\mathcal{A}$
and the affine maps $\tilde{e}^{0}:\mathcal{A}\rightarrow I\!\!R$, $\tilde {e%
}^{0}(z)=1$ and $\tilde{e}^{i}:\mathcal{A}\rightarrow I\!\!R$, $\tilde {e}%
^{i}(z)=z^{i}$, $(\forall)i=\overline{1,n}$. It is easy to see that $%
\mathcal{R}^{\dagger}=\{\tilde{e}^{0},\tilde{e}^{1},\ldots,\tilde{e}%
^{n}\}\subset\mathcal{A}^{\dagger}$ is a base. Let us consider the linear
maps $j:I\!\!R\rightarrow\mathcal{A}^{\dagger}$ and $\pi:\mathcal{A}%
^{\dagger }\rightarrow V^{\ast}$ defined using the bases by $%
j:I\!\!R\rightarrow \mathcal{A}^{\dagger}$, $j(1)=\tilde{e}^{0}$, and $\pi:%
\mathcal{A}^{\dagger }\rightarrow V^{\ast}$, $\pi(\tilde{e}^{0})=0$, $\pi(%
\tilde{e}^{i})=\bar {e}^{i}$, $i=\overline{1,n}$, where $\mathcal{B}%
^{\ast}=\{\bar{e}^{i}\}_{i=\overline{1,n}}\subset V^{\ast}$ is the dual base
of $\mathcal{B}$. It is easy to see that the definitions of $j$ and $\pi$ do
not depend on the frame $\mathcal{R}$ and there is a short exact sequence of
vector spaces having the form $0\rightarrow I\!\!R\overset{j}{\rightarrow}{}%
\mathcal{A}^{\dagger}\overset{\pi}{\rightarrow}{}V^{\ast}\rightarrow0$.
Notice that $\pi$ is also the projection of an affine bundle with the fibre
modeled on $I\!\!R$.

If $\mathcal{R}=(o,\mathcal{B})$ and $\mathcal{R}^{\prime}=(o^{\prime },%
\mathcal{B}^{\prime})$ are two affine frames, then $z^{i}=$ $%
a_{i^{\prime}}^{i}z^{i^{\prime}}+a^{i}$, thus $\tilde{e}^{i}=a_{i^{%
\prime}}^{i}\tilde {e}^{i^{\prime}}+a^{i}\tilde{e}^{0}$. Considering the
bases $\mathcal{R}^{\dagger}$, $\left( \mathcal{R}^{\prime}\right)
^{\dagger}\subset \mathcal{A}^{\dagger}$, then $\xi\in\mathcal{A}^{\dagger}$
has the forms $\xi=\omega\tilde{e}^{0}+\Omega_{i}\tilde{e}%
^{i}=\omega^{\prime}\tilde{e}^{0}+\Omega_{i^{\prime}}\tilde{e}^{i^{\prime}}$
and the following formulas hold: 
\begin{align*}
\Omega_{i^{\prime}} & =a_{i^{\prime}}^{i}\Omega_{i}, \\
\omega^{\prime} & =\omega+\Omega_{i}a^{i}.
\end{align*}

Let us consider now lagrangians and hamiltonians on a real vector space $V$.

A \emph{lagrangian} (a \emph{hamiltonian}) on $V$ is a differentiable map $%
L:V\backslash V\mathcal{_{0}\rightarrow} I\!\!R$ (respectively $H:V^{\ast
}\backslash W_{0}\mathcal{\rightarrow} I\!\!R$), where $V_{0}\subset V$
(respectively $W_{0}\subset V^{\ast}$) is a closed subset (for example an
affine subspace). Differential of $L$ (or $H$) defines the Legendre map
(Legendre$^{\ast}$ map respectively). If the hessian of $L$ (respectively $H 
$) is non-degenerated in every point, then one say that $L$ (respectively $H$%
) is \emph{regular}. In particular, if the hessian of $L$ (respectively $H$)
is strict positively defined, then $L$ (respectively $H$) is regular.
Regular lagrangians and hamiltonians are hyperregular provided that Legendre
maps are diffeomorphisms on their images; they are related by a duality
given by Legendre maps, using the relation $L(z^{i})+H(\Omega_{i})=z^{i}%
\Omega_{i}$.

Let us consider now lagrangians and hamiltonians on an affine space $%
\mathcal{A} $.

A \emph{lagrangian} on $\mathcal{A}$ is a differentiable map $L:\mathcal{A}%
\backslash\mathcal{A}_{0}\mathcal{\rightarrow} I\!\!R$, where $\mathcal{A}%
_{0}\subset\mathcal{A}$ is a closed subset (for example an affine subspace).
If the hessian of $L$ is non-degenerated, then we say that $L$ is \emph{%
regular}. In particular, if the hessian of $L$ is strict positively defined,
then $L$ is regular. The Legendre map $\mathcal{L}:\mathcal{A}\backslash%
\mathcal{A_{0}}\rightarrow V^{\ast}$ is defined also by the differential of $%
L$. If $\mathcal{L}$ is a diffeomorphism on its image, we say that $L$ is
hyperregular; in this case $L$ can be related with a hamiltonian $H$ on $%
V^{\ast}$ by mean of a point $z_{0}\in\mathcal{A}\backslash \mathcal{A}_{0}$%
, using the relation $L(z^{i})+H(\Omega_{i})=(z^{i}-z_{0}^{i})\Omega_{i}$.
We say also that $H$ is \emph{hyperregular}. The consideration of $z_{0}$
gives a $H$ (called a \emph{vectorial hamiltonian}), thus the duality is not
intrinsic. We see below that it is possible to have an intrinsic duality.

An \emph{affine hamiltonian} on the real affine space $\mathcal{A}$ is a
section of an open fibered submanifold of the affine bundle $\mathcal{A}%
^{\dagger}\overset{\pi}{\rightarrow}{}V^{\ast}$. It is defined by a
differentiable map $h:V^{\ast}\backslash W_{0}\rightarrow\mathcal{A}%
^{\dagger }$, such that $\pi\circ h=1_{V^{\ast}\backslash W_{0}}$, where $%
W_{0}\subset V^{\ast}$ is a closed subset. Using an affine frame $(o,%
\mathcal{B})$, then $h$ has the form $h(\Omega_{i})=(\Omega_{i},H_{0}(%
\Omega_{i}))$. If an other affine frame $(o^{\prime},\mathcal{B}^{\prime})$
is considered, then $H_{0}^{\prime}(\Omega_{i^{\prime}})=H_{0}(\Omega_{i})+%
\Omega_{i}a^{i}$. It follows that ${{\displaystyle
{\frac{\partial^{2}H_{0}^{\prime}}{\partial\Omega^{i^{\prime}}\partial
\Omega^{j^{\prime}}}}}}=a_{i^{\prime}}^{i}a_{j^{\prime}}^{j}{{\displaystyle
{\frac{\partial^{2}H_{0}}{\partial\Omega^{i}\partial\Omega^{j}}}}}$, thus
the local functions $H_{0}^{\prime}$ and $H_{0}$ have the same hessian,
which depend only on $h$. We call the hessian of $H_{0}^{\prime}$ and $H_{0}$
as the \emph{hessian} of $h$ and we say that $h$ is \emph{regular} if its
hessian is non-degenerate.

Let $h:V^{\ast}\backslash W_{0}\rightarrow\mathcal{A}^{\dagger}$ be an
affine hamiltonian and consider a point $z_{0}\in\mathcal{A}$. The fact that 
$H_{0}(\Omega_{i})-\Omega_{i}z_{0}^{i}=H_{0}^{\prime}(\Omega_{i^{\prime}})-%
\Omega_{i}\cdot(z_{0}^{i}+a^{i})=$ $H_{0}^{\prime}(\Omega_{i^{\prime}})-%
\Omega_{i}a_{i^{\prime}}^{i}z_{0}^{i^{\prime}}=$ $H_{0}^{\prime}(\Omega_{i^{%
\prime}})-\Omega_{i^{\prime}}z_{0}^{i^{\prime}}$ implies that $%
H(\Omega_{i})=H_{0}(\Omega_{i})-\Omega_{i}z_{0}^{i}$ defines a vectorial
hamiltonian which is regular iff $h$ is regular. Conversely, if $H:V^{\ast
}\backslash W_{0}\rightarrow I\!\!R$ is a vectorial hamiltonian and $z_{0}\in%
\mathcal{A}$ is a point, then denoting $H_{0}(\Omega_{i})=H(\Omega
_{i})+\Omega_{i}z_{0}^{i}$, the map $h:V^{\ast}\backslash W_{0}\rightarrow 
\mathcal{A}^{\dagger}$ given by $h(\Omega_{i})=(\Omega_{i},H_{0}(\Omega_{i}))
$ defines an affine hamiltonian. We say that $h$ is \emph{hyperregular} if $H
$ is hyperregular; it is clear that the definition does not depend on the
point $z_{0}$.

Thus the vectorial and affine hamiltonians are related as follows.

\begin{pr}
\label{praffham1}If $z_{0}\in\mathcal{A}$ is a given point and $W_{0}\subset
V^{\ast}$ is a closed subset, then there is a bijective correspondence
between affine hamiltonians and vectorial hamiltonians on $%
V^{\ast}\backslash W_{0}$.
\end{pr}

Notice that the correspondence defined above depends on the given point $%
z_{0}\in\mathcal{A}$.

A given point $z_{0}\in\mathcal{A}$ and the canonical duality $%
\varphi:V\times V^{\ast}\rightarrow I\!\!R$, define together the \emph{%
Liouville map} $C_{z_{0}}:\mathcal{A}\times V^{\ast}\rightarrow I\!\!R$,
given by the formula $C_{z_{0}}(z,\Omega)=\varphi(z-z_{0},\Omega)$, where $%
z-z_{0}$ denotes the vector $\overline{z_{0}z}$.

\begin{pr}
\label{praffham2}Let $L:\mathcal{A}\backslash\mathcal{A}_{0}\mathcal{%
\rightarrow} I\!\!R$ be a hyperregular lagrangian on the real affine space $%
\mathcal{A}$ and $\mathcal{L}:\mathcal{A}\backslash\mathcal{A}%
_{0}\rightarrow V^{\ast}\backslash W_{0}$ be the Legendre map. Then for
every point $z_{0}\in\mathcal{A}$, the map $H:V^{\ast}\backslash
W_{0}\rightarrow I\!\!R$, $H(\Omega)=C_{z_{0}}(\mathcal{L}%
^{-1}(\Omega),\Omega)-L(\mathcal{L}^{-1}(\Omega))$ is a hamiltonian on $%
V^{\ast}\backslash W_{0}$ and the affine hamiltonian $h:V^{\ast}\backslash
W_{0}\rightarrow\mathcal{A}^{\dagger}$ corresponding to the point $z_{0}$
(according to Proposition \ref{praffham1}) does not depend on the point $%
z_{0}$, depending only on the lagrangian $L$.
\end{pr}

\emph{Proof.} Using coordinates, the link between $L$ and $H$ is $%
L(z^{i})+H(\Omega_{i})=(z^{i}-z_{0}^{i})\Omega_{i}$, where $\mathcal{L}%
^{-1}(\Omega))=z^{i}\bar{e}_{i}$. It is easy to check (classical) that $H$
is a hamiltonian. The affine hamiltonian corresponding to the point $z_{0}$
according to Proposition \ref{praffham1} has the form $(\Omega_{i})\overset{h%
}{\rightarrow}{}(\Omega_{i},H_{0}(\Omega_{i}))$, where $H_{0}(\Omega
_{i})=H(\Omega_{i})+z_{0}^{i}\Omega_{i}=z^{i}\Omega_{i}-L(z^{i})$, thus the
conclusion follows. q.e.d.

A converse correspondence may be performed as follows.

\begin{pr}
\label{praffham3}Let $h:V^{\ast}\backslash W_{0}\rightarrow\mathcal{A}%
^{\dagger}$ be a hyperregular affine hamiltonian on the real affine space $%
\mathcal{A}$. Consider a point $z_{0}\in\mathcal{A}$, the hyperregular
vectorial hamiltonian $H:V^{\ast}\backslash W_{0}\rightarrow I\!\!R$
corresponding to the point $z_{0}$ (according to Proposition \ref{praffham1}%
), $\mathcal{H}:V^{\ast}\backslash W_{0}\rightarrow V\backslash W_{1}$ its
Legendre map and $\mathcal{A}_{0}=z_{0}+W_{1}$. Then

\begin{enumerate}
\item The map $\mathcal{H}_{0}:V^{\ast}\backslash W_{0}\rightarrow \mathcal{A%
}\backslash\mathcal{A}_{0}$ given by the formula $\mathcal{H}_{0}(\Omega)=%
\mathcal{H}(\Omega)+z_{0}$ is a diffeomorphism (called the \emph{Legendre map%
} of $h$).

\item The real function $L:\mathcal{A}\backslash\mathcal{A}_{0}\rightarrow
I\!\!R$ given by the formula $L(z)=$ $C_{z_{0}}(z,\mathcal{H}^{-1}(z-z_{0}))-
$ $H(\mathcal{H}^{-1}(z-z_{0}))$ is a hyperregular lagrangian.

\item Both $\mathcal{H}_{0}$ and $L$ do not depend on the point $z_{0}$,
depending only on the affine hamiltonian $h$.
\end{enumerate}
\end{pr}

\emph{Proof.} Using coordinates, $h$ has the form $(\Omega_{i})\overset{h}{%
\rightarrow}{}(\Omega_{i},H_{0}(\Omega_{i}))$ and $H(\Omega_{i})=H_{0}(%
\Omega_{i})-z_{0}^{i}\Omega_{i}$. Thus $\mathcal{H}(\Omega )^{i}={{%
\displaystyle{\frac{\partial H}{\partial\Omega_{i}}}}}=$ ${{\displaystyle{%
\frac{\partial H_{0}}{\partial\Omega_{i}}}}}-z_{0}^{i}$, then 1. follows,
since $h$ is hyperregular. The proof of 2. uses a similar argument as in the
lagrangian case. Using also coordinates, the link between $L$ and $H$ is $%
L(z)+H(\Omega)=(z^{i}-z_{0}^{i})\Omega_{i}$, where $\Omega=\Omega _{i}\bar{e}%
^{i}=\mathcal{H}^{-1}(z-z_{0})$. It is also easy to check (classical) that $L
$ is a lagrangian. If the affine hamiltonian $h$ has the form $%
h(\Omega)=(\Omega_{i}\bar{e}^{i},H_{0}(\Omega_{i}))$, then $H$ has the form $%
H(\Omega)=H_{0}(\Omega)-z_{0}^{i}\Omega_{i}$, where $\mathcal{H}%
^{-1}(z-z_{0})=$ $\Omega_{i}\bar{e}^{i}=$ $\Omega$. Thus $L(z)=$ $%
(z^{i}-z_{0}^{i})\Omega_{i}-$ $H(\Omega)=$ $(z^{i}-z_{0}^{i})\Omega_{i}-$ $%
H_{0}(\Omega)+z_{0}^{i}\Omega_{i}=$ $z^{i}\Omega_{i}+$ $H_{0}(\Omega)$, thus
2. follows. Using coordinates as above, it follows that the affine
coordinates of $\mathcal{H}_{0}(\Omega)$ are $\left( {{\displaystyle{\frac{%
\partial H_{0}}{\partial\Omega_{i}}}}}\right) $, thus $\mathcal{H}_{0}$
depends only on $H_{0}$ and implicitly on $h$. Taking the coordinates $%
(z^{i})$ of $z\in\mathcal{A}\backslash\mathcal{A}_{0}$ in the form $z^{i}={{%
\displaystyle
{\frac{\partial H_{0}}{\partial\Omega_{i}}}}}$ and denoting, as before, $%
\mathcal{H}^{-1}(z-z_{0})=\Omega_{i}\bar{e}^{i}=\Omega$, we have $\mathcal{H}%
(\Omega)=z-z_{0}$. Using also 2., we have $\mathcal{H}_{0}(\Omega)=\mathcal{H%
}(\Omega)+z_{0}=z$, thus $\Omega=\mathcal{H}_{0}^{-1}(z)$. Since $%
L(z)=z^{i}\Omega_{i}+$ $H_{0}(\Omega)$, the conclusion follows. q.e.d.

\section{An inductive construction of the higher acceleration bundles\label%
{sect1}}

An inductive definition of the higher order spaces $T^{k}M$ are performed,
for example in \cite[Chap. 3, Sect. 1.4]{AG1}, where the notation $J^{k}$ is
used for $T^{k}M$. According to this, using our notations, $T^{0}M=M$, $%
T^{1}M=TM$, $\pi_{1}:T^{1}M\rightarrow T^{0}M$ is the canonical projection
and for $k\geq1$, $T^{k+1}M$ is the affine subbundle of the tangent bundle $%
TT^{k}M$ of vectors $\xi\in T_{x}T^{k}M$ such that considering the
differential $\pi_{k\ast}:T_{x}T^{k}M\rightarrow T_{\pi(x)}T^{k-1}M$ of the
projection $\pi_{k}:T^{k}M\rightarrow T^{k-1}M$, then $\pi_{k\ast}(\xi)=x$
and $\pi _{k+1}:T^{k+1}M\rightarrow T^{k}M$ is induced by the canonical
projection $TT^{k}M\rightarrow T^{k}M$. Notice that it follows a inclusion
map $h_{k}:T^{k}M\rightarrow TT^{k-1}M$, which is an affine bundle map. The
definition of $T^{k}M$ is very simple and has a geometric description, but
is difficult to be used, for example in local coordinates. We give below a
different construction, using a vector pseudofield which comes from the
Liouville vector field.

Let $M$ be a manifold of dimension $m$ and $\tau M=(TM,\pi,M)$ its tangent
bundle. Considering an atlas of $M$, we denote by $(x^{i})$ the coordinates
on an arbitrary domain $U\subset M$ and by $(x^{i},y^{j})$ the coordinates
on the domain $\pi^{-1}(U)\subset TM$ ($i,j=\overline{1,m}$). On the
intersection of two open domains of coordinates on $TM$, the coordinates
change according to the rules 
\begin{equation*}
x^{i\prime}=x^{i\prime}(x^{i}),\, y^{i^{\prime}}={{{\displaystyle
{\frac{\partial x^{i^{\prime}} }{\partial x^{i}}}}}}y^{i}. 
\end{equation*}

A surjective submersion $E\overset{\pi}{\rightarrow}M$ is usually called a 
\emph{fibered manifold}. An \emph{affine bundle} $E\overset{\pi}{\rightarrow 
}M$ is a fibered manifold which the change rules of the local coordinates on 
$E$ have the form 
\begin{equation}
\bar{x}^{i}=\bar{x}^{i}(x^{j}),\bar{y}^{\alpha}=g_{\beta}^{\alpha}(x^{j})y^{%
\beta}+v^{\alpha}(x^{j}).   \label{coq}
\end{equation}
An \emph{affine section} in the bundle $E$ is a differentiable map $M\overset%
{s}{\rightarrow}E$ such that $\pi\circ s=id_{M}$ and its local components
change according to the rule $\bar{s}^{\alpha}(\bar{x}^{i})=g_{\beta}^{%
\alpha}(x^{j})\bar{s}^{\beta}(x^{j})+v^{\alpha}(x^{j})$. The set of affine
sections is denoted by $\Gamma(E)$ and it is an affine module over $\mathcal{%
F}(M)$, i.e. for every $f_{1},\ldots,f_{p}\in\mathcal{F}(M)$ such that $%
f_{1}+\cdots+f_{p}=1$ and $s_{1},\ldots,s_{p}\in\Gamma(E)$, then $%
f_{1}s_{1}+\cdots+f_{p}s_{p}\in\Gamma(E)$, where the affine combination is
taken in every point of the base. Using a partition of unity on the base $M$
one can be easily proved that every affine bundle allows an affine section.

A vector bundle $\bar{E}\overset{\tilde{\pi}}{\rightarrow}M$ can be
canonically associated with the affine bundle $E\overset{\pi}{\rightarrow}M$%
. More precisely, using local coordinates, the coordinates change on $\bar{E}
$ following the rules $\bar{x}^{i}=\bar{x}^{i}(x^{j})$, $\bar{z}^{\alpha
}=g_{\beta}^{\alpha}(x^{j})z^{\beta}$, when the coordinates on $E $ change
according the formulas (\ref{coq}).

Every vector bundle is an affine bundle, called a \emph{central affine bundle%
}. In this case $v^{\alpha}(x^{j})=0$.

Let $E\overset{\pi}{\rightarrow}M$ be a fibered manifold. Consider the
subalgebra $\mathcal{F}_{0}=\{\pi^{\ast}f|$ $f\in\mathcal{F}(M)\}\subset 
\mathcal{F}(E)$ of projectable functions. A \emph{derivation} on the fibered
manifold is an $I\!\!R$-linear map $\Gamma:\mathcal{F}_{0}\rightarrow 
\mathcal{F}(E)$ such that $\Gamma(f\cdot g)=$ $\Gamma(f)\cdot g+f\cdot
\Gamma(g)$, $(\forall)f,g\in\mathcal{F}_{0}$. Let us associate with every
domain of adapted coordinates $\pi^{-1}(U)\subset E$ the vector field $%
\Gamma_{U}=\Gamma(x^{i}){{{\displaystyle{\frac{\partial}{\partial x^{i}}}}}}$%
, which we call the \emph{action} of $\Gamma$ on $\mathcal{F}(\pi^{-1}(U))$.
We call this association a \emph{vector pseudofield} on $E$. For example $%
\Gamma^{(1)}=y^{i}{{{\displaystyle{\frac{\partial}{\partial x^{i}}}}}}$
defines a vector pseudofield on the vector bundle $TM\overset{\pi}{%
\rightarrow}{}M $.

The tangent bundle $TM$ is a vector bundle, but, for $k\geq2$, the $k$%
-accelerations bundles $T^{k}(M)$ are affine bundles over $T^{k-1}(M)$, that
can be defined inductively as follows.

Let us denote $M=T^{0}M$, $\pi=\pi_{1}$ and $TM=T^{1}M$ and consider the
vector pseudofield $\Gamma^{(1)}=y^{i}{{{\displaystyle{\frac{\partial }{%
\partial x^{i}}}}}}$ on the vector bundle $T^{1}M\overset{\pi^{1}}{%
\rightarrow}{}T^{0}M$. Let us suppose that the vector pseudofields $%
\Gamma^{(r)}$ on $T^{r}M$ and the $r$-acceleration bundles $T^{r}M$ have
been defined for $1\leq r\leq k-1$, as affine bundles over $T^{r-1}M$. Then $%
T^{k}M$ is defined according to the change rule of the local coordinates
given by $ky^{(k)i^{\prime}}=k{{{\displaystyle{\frac{\partial x^{i^{\prime}} 
}{\partial x^{i}}}}}}y^{(k)i}+\Gamma_{U}^{(k-1)}(y^{(k-1)i^{\prime}})$ and
the vector pseudofield $\Gamma_{U}^{(k)}=\Gamma_{U}^{(k-1)}+ky^{(k)i}{{{%
\displaystyle{\frac{\partial}{\partial y^{(k-1)i}}}}}}$, where $\Gamma
_{U}^{(k-1)}$ is considered as a (local) vector field on $T^{k}M$ and $U$ is
the domain which corresponds to the coordinates $(x^{i})$. Notice that $%
\Gamma_{U}^{(k)}=y^{(1)i}{{{\displaystyle{\frac{\partial}{\partial x^{i}}}}}}%
+2y^{(2)i}{{{\displaystyle{\frac{\partial}{\partial y^{(1)i}}}}}}%
+\cdots+ky^{(k)i}{{{\displaystyle{\frac{\partial}{\partial y^{(k-1)i}}}}}} $
and on the intersection of two domains corresponding to $U$ and $U^{\prime}$%
, we have $\Gamma_{U^{\prime}}^{(k)}=\Gamma_{U}^{(k)}-%
\Gamma_{U}^{(k)}(y^{(k)i^{\prime}}){{{\displaystyle{\frac{\partial}{\partial
y^{(k)i^{\prime}}}}}}}$.

\begin{pr}
\label{prdefaff}The fibered manifold $(T^{k}M,\pi_{k},T^{k-1}M)$ is an
affine bundle, for $k\geq2$.
\end{pr}

Notice that the coordinates $y^{(p)i}$ are in accordance with those
systematically used by R. Miron in \cite{Mi-Lk}-\cite{MiAt2}.

The vector bundle canonically associated with the affine bundle $(T^{k}M$, $%
\pi_{k},T^{k-1}M)$ is the vector bundle $q_{k-1}^{\ast}TM$, where $%
q_{k-1}:T^{k-1}M\rightarrow M$ is $q_{k-1}=\pi_{1}\circ\pi_{2}\circ
\cdots\circ\pi_{k-1}$. The fibered manifold $(T^{k}M,q_{k},M)$ is
systematically used in \cite{MiAt2, Mi-Lk} in the study of the geometrical
objects of order $k$ on $M$, in particular the Lagrangians of order $k$ on $%
M $. The total space of the dual $q_{k-1}^{\ast}T^{\ast}M$ of the vector
bundle $q_{k-1}^{\ast}TM$ is also the total space of the fibered manifold $%
(T^{k-1}M\times_{M}T^{\ast}M,r_{k},M)$ and it is used in \cite{MiHS, Mi-ha-B}
in the study of the dual geometrical objects of order $k$ on $M$, in
particular the hamiltonians of order $k$ on $M$. In the sequel we denote $%
q_{k-1}^{\ast}T^{\ast}M=T^{k\ast}M$, considered as a vector bundle over $%
T^{k-1}M$.

The tensors defined on the fibers of the vertical vector bundle $%
V_{k-1}^{k}M\rightarrow T^{k}M$ of the affine bundle $(T^{k}M$, $%
\pi_{k},T^{k-1}M)$, or on the fibers of an open fibered submanifold of $%
V_{k-1}^{k}M\rightarrow T^{k}M$, are called \emph{d-tensors of order }$k$ on 
$M$.

Considering local coordinates: $(x^{i}$, $y^{(1)i}$,$\ldots$, $y^{(k-1)i})$
on $T^{k-1}M$ and $(x^{i}$, $y^{(1)i}$,$\ldots$, $y^{(k-1)i}$, $%
p_{(0)i},\ldots,p_{(k-1)i})$ on $T^{\ast}T^{k-1}M$, they change according to
the rules: 
\begin{equation}
\begin{array}{crcr}
y^{(1)i^{\prime}}= & y^{(1)i}{{{\displaystyle{\frac{\partial x^{i^{\prime}}}{%
\partial x^{i}}}}}\qquad} &  &  \\ 
2y^{(2)i^{\prime}}= & y^{(1)i}{{{\displaystyle{\frac{\partial
y^{(1)i^{\prime }}}{\partial x^{i}}}}}}+ & 2y^{(2)i}{{{\displaystyle{\frac{%
\partial y^{(1)i^{\prime}}}{\partial y^{(1)i}}}}}}, &  \\ 
\vdots &  &  &  \\ 
(k-1)y^{(k-1)i}= & y^{(1)i}{{{\displaystyle{\frac{\partial
y^{(k-2)i^{\prime}}}{\partial x^{i}}}}}}+ & \cdots & +(k-1)y^{(k-1)i}{{{%
\displaystyle {\frac{\partial y^{(k-2)i^{\prime}}}{\partial y^{(k-2)i}}}}}}.%
\end{array}
\label{0eqchy}
\end{equation}
and: 
\begin{equation}
\begin{array}{crrr}
p_{(0)i}= & {{{\displaystyle{\frac{\partial x^{i^{\prime}}}{\partial x^{i}}}}%
}}p_{(0)i^{\prime}}+ & {{{\displaystyle{\frac{\partial y^{(1)i^{\prime}}}{%
\partial x^{i}}}}}}p_{(1)i^{\prime}}+\cdots & +{{{\displaystyle {\frac{%
\partial y^{(k-1)i^{\prime}}}{\partial x^{i}}}}}}p_{(k-1)i^{\prime}}, \\ 
p_{(2)i}= &  & {{{\displaystyle{\frac{\partial y^{(1)i^{\prime}}}{\partial
y^{(1)i}}}}}}p_{(1)i^{\prime}}+\cdots & +{{{\displaystyle{\frac{\partial
y^{(k-1)i^{\prime}}}{\partial y^{(1)i}}}}}}p_{(k-1)i^{\prime}}, \\ 
\vdots &  &  &  \\ 
p_{(k-1)i}= &  &  & {{{\displaystyle{\frac{\partial y^{(k-1)i^{\prime}}}{%
\partial y^{(k-1)i}}}}}}p_{(k-1)i^{\prime}}%
\end{array}
\label{0eqchpi}
\end{equation}
respectively.

From formula (\ref{0eqchpi}) it follows that there is a canonical projection
of an affine bundle 
\begin{equation}
\Pi^{\prime}:T^{\ast}T^{k-1}M\rightarrow T^{k\ast}M,   \label{eqdefcpr}
\end{equation}
where $(x^{i}$, $y^{(1)i}$,$\ldots$, $y^{(k-1)i}$, $p_{(0)i},\ldots
,p_{(k-1)i})\overset{\Pi^{\prime}}{\rightarrow}{}(x^{i}$, $y^{(1)i}$,$\ldots$%
, $y^{(k-1)i},p_{(k-1)i})$.

\section{The variational problem for affine Hamiltonians of order $k$}

In this section we define affine hamiltonians of order $k\geq2$ on a
manifold $M$ and study the variational problem of their integral action on
curves on $T^{\ast}M$. We define also the energy of an affine hamiltonian of
order $k$ as a function on $T^{\ast}T^{k-1}M$ and we prove that the integral
curves of its Hamilton vector field are natural liftings of solutions of the
variational problem of the integral action. The case $k=1$ is that studied
in the classical mechanics, the affine situation occurs only for $k\geq2$.
In that follows we suppose that $k\geq2$, if any other assumption is made.

Let us consider the affine bundle $T^{k}M\overset{\pi_{k}}{\rightarrow}%
{}T^{k-1}M$ and $u\in T^{k-1}M$. The fiber $T_{u}^{k}M=\pi_{k}^{-1}(u)%
\subset T^{k}M$ is a real affine space, modelled on the real vector space $%
T_{\pi (u)}M$. The \emph{vectorial dual} of the affine space $T_{u}^{k}M$ is 
$T_{u}^{k\dagger}M=Aff(T_{u}^{k}M,I\!\!R)$, where $Aff$ denotes affine
morphisms. Denoting by $T^{k\dagger}M=%
\mathrel{\mathop{\cup
}\limits_{u\in T^{k-1}M}}{}T_{u}^{k\dagger}M$ and $\pi^{\dagger}:T^{k\dagger
}M\rightarrow T^{k-1}M$ the canonical projection, then it is clear that $%
(T^{k\dagger}M,\pi^{\dagger},T^{k-1}M)$ is a vector bundle. There is a
canonical vector bundle morphism of vector bundles over $T^{k-1}M$, $%
\Pi:T^{k\dagger}M\rightarrow T^{k\ast}M$. This projection is also the
canonical projection of an affine bundle with the fiber $I\!\!R$. An \emph{%
affine hamiltonian} of order $k$ on $M$ is a pointed section $h:\widetilde{%
T^{k\ast}M}\rightarrow\widetilde{T^{k\dagger}M}$ of this affine bundle
(possibly continuous on the image of the null section), or a section of an
open fibered submanifold of $\Pi$. Thus an affine hamiltonian of order $k$
on $M$ is given by a differentiable map $h:\widetilde{T^{k\ast}M}\rightarrow%
\widetilde{T^{k\dagger}M}$, such that $\Pi\circ h=1_{\widetilde {T^{k\ast}M}}
$.

We consider some local coordinates $(x^{i})$ on $M$, $(x^{i},$ $%
y^{(1)i},\ldots,$ $y^{(k-1)i})$ on $T^{k-1}M$, and $(x^{i},y^{(1)i},\ldots
,y^{(k-1)i},p_{i},T)$ on $T^{k\dagger}M$. Then the coordinates $p_{i}$ and $T
$ change according to the rules $p_{i^{\prime}}={{{\displaystyle{\frac{%
\partial x^{i}}{\partial x^{i^{\prime}}}}}}}p_{i}$ and $T^{\prime}=T+$ ${{{%
\displaystyle{\frac{1}{k}}}}}\Gamma_{U}^{(k-1)}(y^{(k-1)i^{\prime}}){{{%
\displaystyle{\frac{\partial x^{i}}{\partial x^{i^{\prime}}}}}}}p_{i}$
respectively. The vector bundle morphism $\Pi$ is given in local coordinates
by $(x^{i},y^{(1)i},\ldots,y^{(k-1)i},p_{i},T)\overset{\Pi}{\rightarrow}{}$ $%
(x^{i},y^{(1)i},\ldots,y^{(k-1)i},p_{i})$. Thus the local function $H_{0}$
changes according to the rules 
\begin{align}
H_{0}^{\prime}(x^{i^{\prime}},y^{(1)i^{\prime}},\ldots,y^{(k-1)i^{%
\prime}},p_{i^{\prime}}) & =H_{0}(x^{i},y^{(1)i},\ldots,y^{(k-1)i},p_{i})
\label{eqchh0} \\
& +{{{\displaystyle{\frac{1}{k}}}}}\Gamma_{U}^{(k-1)}(y^{(k-1)i^{\prime}}){{{%
\displaystyle{\frac{\partial x^{i}}{\partial x^{i^{\prime}}}}}}}p_{i}. 
\notag
\end{align}
Notice that the corresponding map $h:\widetilde{T^{k\ast}M}\rightarrow 
\widetilde{T^{k\dagger}M}$ has the local form $h(x^{i},y^{(1)i},\ldots
,y^{(k-1)i},p_{i})=$ $(x^{i},y^{(1)i},%
\ldots,y^{(k-1)i},p_{i},H_{0}(x^{i},y^{(1)i},\ldots,y^{(k-1)i},p_{i}))$.

It is easy to see that ${{{\displaystyle{\frac{\partial H_{0}^{\prime}}{%
\partial p_{i^{\prime}}}}}}}={{{\displaystyle{\frac{\partial x^{i^{\prime}}}{%
\partial x^{i}}}}}}{{{\displaystyle{\frac{\partial H_{0}}{\partial p_{i}}}}}}%
+{{{\displaystyle{\frac{1}{k}}}}}\Gamma_{U}^{(k-1)}(y^{(k-1)i^{\prime}})$.
Thus there is a map $\mathcal{H}:T^{k\ast}M\rightarrow T^{k}M$, which we
call the \emph{Legendre}$^{\ast}$\emph{\ map}, given in local coordinates by 
$(x^{i},y^{(1)i},\ldots,y^{(k-1)i},$ $p_{i})\overset{\mathcal{H}}{%
\rightarrow }{}(x^{i},y^{(1)i},\ldots,y^{(k-1)i},\mathcal{H}^{i})$, where $%
\mathcal{H}^{i}(x^{i},y^{(1)i},\ldots,y^{(k-1)i},$ $p_{i})=$ ${{{%
\displaystyle
{\frac{\partial H_{0}}{\partial p_{i}}}}}}(x^{i},y^{(1)i},\ldots
,y^{(k-1)i},p_{i})$. We say also that $h$ is \emph{hyperregular} if $%
\mathcal{H}$ is a global diffeomorphism. Since ${{{\displaystyle
{\frac{\partial^{2}H_{0}^{\prime}}{\partial p_{i^{\prime}}\partial
p_{j^{\prime}}}}}}}={{{\displaystyle{\frac{\partial x^{i^{\prime}}}{\partial
x^{i}}}}}}{{{\displaystyle{\frac{\partial x^{j^{\prime}}}{\partial x^{j}}}}}}%
{{{\displaystyle{\frac{\partial^{2}H_{0}}{\partial p_{i}\partial p_{j}}}}}}$%
, it follows that $h^{ij}={{{\displaystyle{\frac{\partial^{2}H_{0}}{\partial
p_{i}\partial p_{j}}}}}}$ is a symmetric $2$-contravariant d-tensor, which
is non-degenerate if $h$ is hyperregular.

For any curve $\gamma:[0,1]\rightarrow T^{\ast}M$ , we define the \emph{%
integral action} of $h$ along the curve $\gamma$ by the formula 
\begin{equation}
I(\gamma)=\int_{0}^{1}\left[ p_{i}{{\displaystyle{\frac{1}{\left( k-1\right)
!}}}}{{\displaystyle{\frac{d^{k}x^{i}}{dt^{k}}}}}-kH_{0}\left( x^{i},{{%
\displaystyle{\frac{dx^{i}}{dt}}}},\ldots,{{\displaystyle{\frac {1}{(k-1)!}}}%
}{{\displaystyle{\frac{d^{k-1}x^{i}}{dt^{k-1}}}}},p_{i}\right) \right] dt, 
\label{0eqig}
\end{equation}
where $\gamma$ has the local form $t\overset{\gamma}{\rightarrow}%
{}(x^{i}(t),p_{i}(t))$. The critical condition (or Fermat condition in the
case of an extremum) can be used for the integral action using similar
arguments as in \cite[Chapter 5]{Mi-ha-B}, where it is used in the case of
extremun and it is called as \emph{extremum condition}; we sketch below
these arguments. The lift $\check{\gamma}:[0,1]\rightarrow T^{k\ast}M$ of $%
\gamma$ has the local form. $t\rightarrow\left( x^{i}(t),{{\displaystyle{%
\frac{dx^{i}}{dt}}}}(t),\right. $ $\left. \ldots,{{\displaystyle{\frac{1}{%
(k-1)!}}}}{{\displaystyle{\frac{d^{k-1}x^{i}}{dt^{k-1}}}}}%
(t),p_{i}(t)\right) $. We consider a vector field $V^{i}$ and a covector
field $\eta_{i}$ along the curve $\gamma$ having the properties that $%
V^{i}(0)=$ $V^{i}(1)=$ ${{\displaystyle{\frac{d^{\alpha}V^{i}}{dt^{\alpha}}}}%
}(0)=$ ${{\displaystyle
{\frac{d^{\alpha}V^{i}}{dt^{\alpha}}}}}(0)=$ $\eta^{i}(0)=$ $\eta^{i}(1)=$ $0
$. For $\varepsilon_{1}$, $\varepsilon_{2}>0$ (small enough), one consider a
variation $\bar{\gamma}(\varepsilon_{1},\varepsilon_{2})$ in the form $%
t\rightarrow\bar{\gamma}^{i}(t)=(x^{i}(t)+\varepsilon_{1}V^{i}(t)$, $%
p_{i}+\varepsilon_{2}\eta^{i}(t))$. Necessary conditions that $I(\gamma)$ be
a critical value of $I(\bar{\gamma}(\varepsilon_{1},\varepsilon_{2}))$ are
given vanishing partial derivatives of $I(\bar{\gamma}(\varepsilon
_{1},\varepsilon_{2}))$ with respect to $\varepsilon_{1}$ and $\varepsilon
_{2}$, in $(\varepsilon_{1}=0$, $\varepsilon_{2}=0)$. This condition gives 
\begin{align*}
& \int_{0}^{1}\left[ p_{i}(t){{\displaystyle{\frac{1}{\left( k-1\right) !}}}}%
{{\displaystyle{\frac{d^{k}V^{i}}{dt^{k}}}}}-\right. \\
& \left. k\left( {{\displaystyle{\frac{\partial H_{0}}{\partial x^{i}}}}}%
V^{i}+{{\displaystyle{\frac{\partial H_{0}}{\partial y^{(1)i}}}}}{{%
\displaystyle{\frac{dV^{i}}{dt}}}}+\cdots+{{\displaystyle{\frac{1}{\left(
k-1\right) !}}}}{{\displaystyle{\frac{\partial H_{0}}{\partial y^{(k-1)i}}}}}%
{{\displaystyle{\frac{d^{k-1}V^{i}}{dt^{k-1}}}}}\right) \right] dt\hbox{=}0
\end{align*}
and 
\begin{equation*}
\int_{0}^{1}\left[ {{\displaystyle{\frac{1}{k!}}}}{{\displaystyle{\frac {%
d^{k}x^{i}}{dt^{k}}}}}-{{\displaystyle{\frac{\partial H_{0}}{\partial
p_{(k-1)i}}}}}\right] \eta_{i}dt=0. 
\end{equation*}

Integrating successively by parts the terms that contain derivatives of $%
V^{i}$, we obtain the following equivalent form of the first integral: 
\begin{equation*}
\int_{0}^{1}\left[ {{\displaystyle{\frac{\left( -1\right) ^{k}}{k!}}}}{{%
\displaystyle{\frac{d^{k}p_{i}}{dt^{k}}}}}-{{\displaystyle{\frac{\partial
H_{0}}{\partial x^{i}}}}}+{{\displaystyle{\frac{d}{dt}}}}{{\displaystyle
{\frac{\partial H_{0}}{\partial y^{(1)i}}}}}-\cdots+{{\displaystyle
{\frac{\left( -1\right) ^{k-1}}{\left( k-1\right) !}}}}{{\displaystyle
{\frac{d^{k-1}}{dt^{k-1}}}}}{{\displaystyle{\frac{\partial H_{0}}{\partial
y^{(k-1)i}}}}}\right] V^{i}dt=0. 
\end{equation*}
Since the curve $\bar{\gamma}$ is arbitrary, we obtain the Hamilton equation
in the condensed form: 
\begin{equation}
\left\{ 
\begin{tabular}{l}
${{\displaystyle{\frac{\left( -1\right) ^{k}}{k!}}}}{{\displaystyle {\frac{%
d^{k}p_{i}}{dt^{k}}}}}-{{\displaystyle{\frac{\partial H_{0}}{\partial x^{i}}}%
}}+{{\displaystyle{\frac{d}{dt}}}}{{\displaystyle{\frac{\partial H_{0}}{%
\partial y^{(1)i}}}}}-\cdots+{{\displaystyle{\frac{\left( -1\right) ^{k-1}}{%
\left( k-1\right) !}}}}{{\displaystyle{\frac{d^{k-1}}{dt^{k-1}}}}}{{%
\displaystyle{\frac{\partial H_{0}}{\partial y^{(k-1)i}}}}}=0,$ \\ 
${{\displaystyle{\frac{1}{k!}}}}{{\displaystyle{\frac{d^{k}x^{i}}{dt^{k}}}}}-%
{{\displaystyle{\frac{\partial H_{0}}{\partial p_{(k-1)i}}}}}=0,$%
\end{tabular}
\right.   \label{0eqhamjaccf}
\end{equation}
where $y^{(1)i}={{\displaystyle{\frac{dx^{i}}{dt}}}}$,$\ldots$, $y^{(k-1)i}={%
{\displaystyle{\frac{1}{\left( k-1\right) !}}}}{{\displaystyle
{\frac{d^{k-1}x^{i}}{dt^{k-1}}}}}$.

Notice that for $k=1$, the formulas (\ref{0eqhamjaccf}) are also valid in
the form: 
\begin{equation*}
\left\{ 
\begin{tabular}{l}
$-{{\displaystyle{\frac{dp_{i}}{dt}}}}-{{\displaystyle{\frac{\partial H_{0}}{%
\partial x^{i}}}}}=0,$ \\ 
${{\displaystyle{\frac{dx^{i}}{dt}}}}-{{\displaystyle{\frac{\partial H_{0}}{%
\partial p_{(0)i}}}}}=0,$%
\end{tabular}
\right. 
\end{equation*}
i.e. the well-known Hamilton equation of the hamiltonian $H_{0}$.

We consider, for an affine $k$-hamiltonian $h$ ($k\geq2$) and the local
domain $U$, the energy $\mathcal{E}_{U}$ given by the local formula: 
\begin{equation}
\mathcal{E}_{U}=p_{(0)i}y^{(1)i}+%
\cdots+(k-1)p_{(k-2)i}y^{(k-1)i}+kH_{0}(x^{i},y^{(1)i},%
\ldots,y^{(k-1)i},p_{(k-1)i}).   \label{eqh0}
\end{equation}

\begin{pr}
The local functions $\mathcal{E}_{U}$ glue together to a global function $%
\mathcal{E}_{0}:T^{\ast}\widetilde{T^{k-1}M}\rightarrow I\!\!R$.
\end{pr}

\emph{Proof.} Using the change rules (\ref{0eqchy}), (\ref{0eqchpi}) and (%
\ref{eqchh0}), we have:

\noindent$p_{(0)i}y^{(1)i}+%
\cdots+(k-1)p_{(k-2)i}y^{(k-1)i}+kH_{0}(x^{i},y^{(1)i},%
\ldots,y^{(k-1)i},p_{(k-1)i})=$

\noindent$\left( {{{\displaystyle{\frac{\partial x^{i^{\prime}} }{\partial
x^{i}}}}}}p_{(0)i^{\prime}}+{{{\displaystyle{\frac{\partial
y^{(1)i^{\prime}} }{\partial x^{i}}}}}}p_{(1)i^{\prime}}+\cdots+{{{%
\displaystyle{\frac{\partial y^{(k-1)i^{\prime}} }{\partial x^{i}}}}}}%
p_{(k-1)i^{\prime}}\right) y^{(1)i}+$

\noindent$2\left( {{{\displaystyle{\frac{\partial y^{(1)i^{\prime}} }{%
\partial y^{(1)i}}}}}}p_{(1)i^{\prime}}+\cdots+{{{\displaystyle
{\frac{\partial y^{(k-1)i^{\prime}} }{\partial y^{(1)i}}}}}}%
p_{(k-1)i^{\prime }}\right) y^{(2)i}+\cdots$

\noindent$+(k-1)\left( {{{\displaystyle{\frac{\partial y^{(k-2)i^{\prime}} }{%
\partial y^{(k-2)i}}}}}}p_{(k-2)i^{\prime}}+{{{\displaystyle{\frac{\partial
y^{(k-1)i^{\prime}} }{\partial y^{(k-2)i}}}}}}p_{(k-1)i^{\prime}}\right)
y^{(k-1)i}+kH_{0}=$

\noindent$p_{(0)i^{\prime}}y^{(1)i^{\prime}}+\cdots+(k-1)p_{(k-2)i^{%
\prime}}y^{(k-1)i^{\prime}}+p_{(k-1)i^{\prime}}%
\Gamma_{U}^{(k-1)}(y^{(k-1)i^{\prime}})+kH_{0}(x^{i},$ $y^{(1)i},\ldots,$ $%
y^{(k-1)i},$ $p_{(k-1)i})$, thus the conclusion follows. q.e.d.

We call the function $\mathcal{E}_{0}$ the \emph{energy} of the affine
hamiltonian $h$.

The manifold $T^{\ast}\widetilde{T^{k-1}M}$ has a canonical symplectic
structure given by $\Omega=dp_{(0)i}\wedge dx^{i}+dp_{(1)i}\wedge
dy^{(1)i}+\cdots+dp_{(k-1)i}\wedge dy^{(k-1)i}$. The hamiltonian vector
field $X_{\mathcal{E}}$ which corresponds to the function $\mathcal{E}$ :$%
T^{\ast }\widetilde{T^{k-1}M}\rightarrow I\!\!R$ is defined according to the
formula $d\mathcal{E}=i_{X_{\mathcal{E}}}\Omega$ and it has the local
expression: 
\begin{align*}
X_{\mathcal{E}} & ={{{\displaystyle{\frac{\partial\mathcal{E} }{\partial
p_{(0)i}}}}}}{{{\displaystyle{\frac{\partial}{\partial x^{i}}}}}}+{{{%
\displaystyle{\frac{\partial\mathcal{E} }{\partial p_{(1)i}}}}}}{{{%
\displaystyle{\frac{\partial}{\partial y^{(1)i}}}}}}+\cdots +{{{\displaystyle%
{\frac{\partial\mathcal{E} }{\partial p_{(k-1)i}}}}}}{{{\displaystyle{\frac{%
\partial}{\partial y^{(k-1)i}}}}}}- \\
& -{{{\displaystyle{\frac{\partial\mathcal{E} }{\partial x^{i}}}}}}{{{%
\displaystyle{\frac{\partial}{\partial p_{(0)i}}}}}}-{{{\displaystyle {\frac{%
\partial\mathcal{E} }{\partial y^{(1)i}}}}}}{{{\displaystyle {\frac{\partial%
}{\partial p_{(1)i}}}}}}-\cdots-{{{\displaystyle{\frac {\partial\mathcal{E} 
}{\partial y^{(k-1)i}}}}}}{{{\displaystyle{\frac {\partial}{\partial
p_{(k-1)i}}}}}}.
\end{align*}

An integral curve of the hamiltonian vector field $X_{\mathcal{E}_{0}}$ is a
solution of the differential equations: 
\begin{equation}
\left\{ 
\begin{array}{l}
{{{\displaystyle{\frac{dx^{i}}{dt}}}}}=y^{(1)i}, \\ 
{{{\displaystyle{\frac{dy^{(\alpha)i}}{dt}}}}}=(\alpha+1)y^{(\alpha
+1)i},\alpha=\overline{1,k-2}, \\ 
{{{\displaystyle{\frac{dy^{(k-1)i}}{dt}}}}}=k{{\displaystyle{\frac{\partial
H_{0}}{\partial p_{(k-1)i}}}}}, \\ 
{{{\displaystyle{\frac{dp_{(0)i}}{dt}}}}}=-k{{{\displaystyle{\frac{\partial
H_{0}}{\partial x^{i}}}}}}, \\ 
{{{\displaystyle{\frac{dp_{(\alpha)i}}{dt}}}}}=-k{{{\displaystyle {\frac{%
\partial H_{0}}{\partial y^{(\alpha)i}}}}}}-\alpha p_{(\alpha -1)i},\alpha=%
\overline{1,k-1},%
\end{array}
\right.   \label{eqjhnewh}
\end{equation}
which we call the \emph{Hamilton} equation (in the vectorial form) of the
affine hamiltonian $h$. Notice that this form and the condensed form (\ref%
{0eqhamjaccf}) are equivalent, giving the Hamilton equation of the affine
hamiltonian $h$.

\begin{theor}
\label{thgeoham}An integral curve $\Gamma$ of the hamiltonian vector field $%
X_{\mathcal{E}_{0}}$ projects on a curve $\gamma$ on $T^{\ast}M$ which is a
solution of the Hamilton equation of the affine hamiltonian $h$ in the form (%
\ref{0eqhamjaccf}), thus $\gamma$ is a critical curve for the action $I$.
\end{theor}

\emph{Proof.} The integral curve $\Gamma$ is a solution of a local
differential equation having the form (\ref{eqjhnewh}). It projects on the
curve $\gamma$ on $T^{\ast}M$ given in local coordinates by $\gamma
(t)=(x^{i}(t),p_{(k-1)i}(t))$. By a straightforward verification, the
differential equations (\ref{eqjhnewh}) implies that the Hamilton equation (%
\ref{0eqhamjaccf}) hold. q.e.d.

Let us regard $\mathcal{E}_{0}$ as a hamiltonian function on $%
T^{\ast}T^{k-1}M$. For any curve $\Gamma:[0,1]\rightarrow T^{\ast}T^{k-1}M$, 
$\Gamma(t)=(x^{i}(t),y^{(1)i}(t),\ldots,$ $y^{(k-1)i}(t),p_{(0)i}(t),%
\ldots,p_{(k-1)i}(t))$ the \emph{integral action} of $\mathcal{E}_{0}$ along
the curve $\Gamma$ is given by the formula 
\begin{align*}
I_{\mathcal{E}_{0}}(\Gamma) & =\int_{0}^{1}\left[ p_{(0)i}{{\displaystyle {%
\frac{dx^{i}}{dt}}}}+p_{(1)i}{{\displaystyle{\frac{dy^{(1)i}}{dt}}}}%
+\cdots+p_{(k-1)i}{{\displaystyle{\frac{dy^{(k-1)i}}{dt}}}}-\right. \\
& \left. \mathcal{E}_{0}\left(
x^{i},y^{(1)i},\ldots,y^{(k-1)i},p_{(0)i},\ldots,p_{(k-1)i}\right) \right]
dt,
\end{align*}
thus 
\begin{align*}
I_{\mathcal{E}_{0}}(\Gamma) & =\int_{0}^{1}\left[ p_{(0)i}\left( {{%
\displaystyle{\frac{dx^{i}}{dt}}}}-y^{(1)i}\right) +p_{(1)i}\left( {{%
\displaystyle{\frac{dy^{(1)i}}{dt}}}}-2y^{(2)i}\right) +\cdots+\right. \\
& p_{(k-2)i}\left( {{\displaystyle{\frac{dy^{(k-2)i}}{dt}}}}%
-(k-1)y^{(k-1)i}\right) + \\
& \left. p_{(k-1)i}{{\displaystyle{\frac{dy^{(k-1)i}}{dt}}}}-kH_{0}\left(
x^{i},y^{(1)i},\ldots,y^{(k-1)i},p_{(k-1)i}\right) \right] dt.
\end{align*}
If the curve $\Gamma$ has the property that 
\begin{equation}
{{\displaystyle{\frac{dx^{i}}{dt}}}}=y^{(1)i},\ldots,{{\displaystyle {\frac{%
dy^{(k-2)i}}{dt}}}}=(k-1)y^{(k-1)i},   \label{eqprol}
\end{equation}
then $I_{\mathcal{E}_{0}}(\Gamma)=I(\gamma)$.

Let us consider a curve $\gamma:[0,1]\rightarrow T^{\ast}M$; we say that a
curve $\Gamma:[0,1]\rightarrow T^{\ast}T^{k-1}M$ is a \emph{lift} of $\gamma$
to $T^{\ast}T^{k-1}M$ if $\check{\gamma}=\Pi^{\prime}\circ\Gamma$, where $%
\check{\gamma}:[0,1]\rightarrow T^{k\ast}M$ is the lift of $\gamma$ to $%
T^{k\ast}M$ and $\Pi^{\prime}$ is the canonical projection defined in (\ref%
{eqdefcpr}), i.e. the diagram 
\begin{equation*}
\begin{array}{crl}
\lbrack0,1] & \overset{\Gamma}{\longrightarrow}{} & T^{\ast}T^{k-1}M \\ 
\downarrow\check{\gamma} & \swarrow & \Pi^{\prime} \\ 
T^{k\ast}M &  & 
\end{array}
\end{equation*}
is commutative.

If $s:U\subset T^{k\ast}M\rightarrow T^{\ast}T^{k-1}M$ is a section of the
fibered manifold $T^{\ast}T^{k-1}M\overset{\Pi^{\prime}}{\rightarrow}%
{}T^{k\ast}M$, where $U$ is an open set of $T^{k\ast}M$ which contains $%
\check{\gamma}([0,1])$, then $\Gamma=s\circ\check{\gamma}$ is a lift of $%
\gamma$ to $T^{\ast}T^{k-1}M$.

We have the following relation between critical curves of actions $I$ and $%
I_{\mathcal{E}_{0}}$.

\begin{theor}
If $\Gamma_{0}$ is a critical curve in $T^{\ast}T^{k-1}M$ for the action $I_{%
\mathcal{E}_{0}}$ and $\gamma_{0}$ is its projection curve in $T^{\ast}M$ ,
then the curve $\gamma_{0}$ is a critical curve for the action $I$ and $I_{%
\mathcal{E}_{0}}(\Gamma_{0})=I(\gamma_{0})$.
\end{theor}

\emph{Proof. }Since $\Gamma_{0}$ is a critical curve in $T^{\ast}T^{k-1}M$
for the action $I_{\mathcal{E}_{0}}$, it is a solution of the vectorial
Hamilton equation in the form (\ref{eqjhnewh}). Using Theorem \ref{thgeoham}%
, it follows that the curve $\gamma_{0}$ in $T^{\ast}M$ is a solution of the
Hamilton equation in the form (\ref{0eqhamjaccf}). q.e.d.

\begin{cor}
If $\Gamma_{0}$ is a critical curve for the action $I_{\mathcal{E}_{0}}$,
then it is a lift to $T^{\ast}T^{k-1}M$ of a critical curve $\gamma_{0}$ for
the action $I$.
\end{cor}

\emph{Proof.} Let $\gamma_{0}$ be the projection of $\Gamma_{0}$ to $T^{\ast
}M$, i.e. $\gamma_{0}$ is obtained as $[0,1]\overset{\Gamma_{0}}{\rightarrow 
}{}T^{\ast}T^{k-1}M\overset{\Pi^{\prime}}{\rightarrow}{}T^{k\ast}M%
\rightarrow T^{\ast}M$. It is clear that $\Gamma_{0}$ is a lift of $%
\gamma_{0}$ and using the above Theorem $\gamma_{0}$ is a critical curve for
the action $I$. q.e.d.

A \emph{vectorial hamiltonian} of order $k$ on $M$ is a function $H:T^{k\ast
}M\rightarrow I\!\!R$, differentiable on $\widetilde{T^{k\ast}M}$ (i.e. $%
T^{k\ast}M$ without the null section). The vectorial hamiltonian is \emph{%
hyperregular} if the vertical Hessian $\left( {{{\displaystyle
{\frac{\partial^{2}H}{\partial p_{i}p_{j}}}}}}\right) $ of $H$ is
non-degenerate and the Legendre$^{\ast}$ map is a diffeomorphism on its
image. In this case the vertical hessian defines a non-degenerate bilinear
form on the fibers of the vertical bundle $V\widetilde{T^{k\ast}M}$. The
vectorial hamiltonian defined here is called simply a hamiltonian in \cite%
{MiHS, Mi-ha-B}. We say that this hamiltonian is \emph{vectorial}, in order
to distinguish it from the \emph{affine hamiltonian} already defined before.

Using Proposition \ref{praffham1}, we obtain the following relation between
affine and vectorial hamiltonians.

\begin{pr}
\label{prcorham}Let $s:T^{k-1}M\rightarrow T^{k}M$ be an affine section.

\begin{enumerate}
\item If $H:T^{k\ast}M\rightarrow I\!\!R$ is a vectorial hamiltonian, then
the local functions $H_{0}=H(x^{i},y^{(1)i},\ldots,y^{(k-1)i},p_{i})+{s}%
^{i}(x^{i},y^{(1)i},\ldots,y^{(k-1)i})p_{i}$ define an affine hamiltonian $h$
of order $k$ on $M$.

\item Conversely, if $h$ is an affine hamiltonian of order $k$ on $M$, then
the local functions $H=H_{0}(x^{i},y^{(1)i},\ldots,y^{(k-1)i},p_{i})-{s}%
^{i}(x^{i},y^{(1)i},\ldots,y^{(k-1)i})p_{i}$ define a vectorial hamiltonian
of order $k$ on $M$.
\end{enumerate}
\end{pr}

Notice that a vectorial hamiltonian $H$ and an affine hamiltonian $h$ define
together an affine section on the affine bundle $\widetilde{T^{k}M}%
\rightarrow\widetilde{T^{k-1}M}$; locally, this is given by $s^{i}={%
\displaystyle{\frac{\partial H_{0} }{\partial p_{i}}}}-{\displaystyle
{\frac{\partial H }{\partial p_{i}}}}$.

Thus an affine section is an essential ingredient in a correspondence
between vectorial and affine hamiltonians.

\section{Zermelo conditions}

The Zermelo conditions in higher order geometry are discussed for
lagrangians in \cite[Ch.8]{Mi-Lk} and for vectorial hamiltonians in \cite[%
Ch.5]{Mi-ha-B}. The integral action considered here for an affine
hamiltonian is completely different from that used in \cite[Ch.5]{Mi-ha-B}
for a vectorial hamiltonian, but the calculation of Zermelo conditions
follows the same ideas.

The Zermelo conditions are necessary conditions imposed to the affine
hamiltonian, in order that the integral action (\ref{0eqig}) do not depend
on the parametrization of an arbitrary curve $\gamma:[0,1]\rightarrow
T^{\ast}M$.

Let $\tilde{t}:[0,1]\rightarrow\lbrack a,b]$ ($\tilde{t}(0)=a$, $\tilde {t}%
(1)=b$), be a diffeomorphism that defines a new parametrization of the curve 
$\gamma$, which has the local form $\gamma(\tilde{t})=(\tilde{x}^{i}(\tilde{t%
}),\tilde{p}_{i}(\tilde{t}))$, $\tilde{t}\in\lbrack a,b]$. The condition
that the integral action (\ref{0eqig}) do not depend on the parametrization
leads to the condition that the integrand terms be the same, i.e. 
\begin{equation}
\begin{tabular}{l}
$\left[ {{\displaystyle{\frac{1 }{\left( k-1\right) !}}}}\tilde{p}_{i}{{%
\displaystyle{\frac{d^{k}\tilde{x}^{i} }{d\tilde{t}^{k}}}}}-kH_{0}\left( 
\tilde{x}^{i},{{\displaystyle{\frac{d\tilde{x}^{i} }{d\tilde {t}}}}},\ldots,{%
{\displaystyle{\frac{1 }{\left( k-1\right) !}}}}{{\displaystyle{\frac{d^{k-1}%
\tilde{x}^{i} }{d\tilde{t}^{k-1}}}}},\tilde {p}_{i}\right) \right] {{%
\displaystyle{\frac{d\tilde{t} }{dt}}}}=$ \\ 
$={{\displaystyle{\frac{1 }{\left( k-1\right) !}}}}p_{i}{{\displaystyle {%
\frac{d^{k}x^{i} }{dt^{k}}}}}-kH_{0}\left( x^{i},{{\displaystyle {\frac{%
dx^{i} }{dt}}}},\ldots,{{\displaystyle{\frac{1 }{\left( k-1\right) !}}}}{{%
\displaystyle{\frac{d^{k-1}x^{i} }{dt^{k-1}}}}},p_{i}\right) .$%
\end{tabular}
\label{0eqzerm}
\end{equation}
The right side of the above equality can be regarded as a function on $t$,
but via $\tilde{t}$. We have ${{\displaystyle{\frac{dx^{i} }{dt}}}}={{%
\displaystyle{\frac{dx^{i} }{d\tilde{t}}}}}{{\displaystyle{\frac{d\tilde{t} 
}{dt}}}}$, ${{\displaystyle{\frac{d^{2}x^{i} }{dt^{2}}}}}={{\displaystyle
{\frac{d^{2}x^{i} }{d\tilde{t}^{2}}}}}\left( {{\displaystyle{\frac{d\tilde{t}
}{dt}}}}\right) ^{2}+$ ${{\displaystyle{\frac{dx^{i} }{d\tilde{t}}}}}{{%
\displaystyle{\frac{d^{2}\tilde{t} }{dt^{2}}}}}$, ${{\displaystyle
{\frac{d^{3}x^{i} }{dt^{3}}}}}={{\displaystyle{\frac{d^{3}x^{i} }{d\tilde {t}%
^{3}}}}}\left( {{\displaystyle{\frac{d\tilde{t} }{dt}}}}\right) ^{3}+3{{%
\displaystyle{\frac{d^{2}x^{i} }{d\tilde{t}^{2}}}}}{{\displaystyle
{\frac{d\tilde{t} }{dt}}}}{{\displaystyle{\frac{d^{2}\tilde{t} }{dt^{2}}}}}+$
${{\displaystyle{\frac{dx^{i} }{d\tilde{t}}}}}{{\displaystyle{\frac {d^{3}%
\tilde{t} }{dt^{3}}}}}$ $\allowbreak$and so on. The left side of the
equality (\ref{0eqzerm}) depends on the diffeomorphism $\tilde{t}$ via the
derivative ${{\displaystyle{\frac{d\tilde{t} }{dt}}}}$, while the right side
depends on the diffeomorphism $\tilde{t}$ via the derivatives ${{%
\displaystyle
{\frac{d\tilde{t} }{dt}}}}$, ${{\displaystyle{\frac{d^{2}\tilde{t} }{dt^{2}}}%
}}$,$\ldots$ , ${{\displaystyle{\frac{d^{k}\tilde{t} }{dt^{k}}}}}$.
Differentiating both sides of the equality (\ref{0eqzerm}) with respect to ${%
{\displaystyle{\frac{d\tilde{t} }{dt}}}}$, then letting $t=\tilde{t}$, we
have 
\begin{equation*}
\begin{tabular}{l}
${{\displaystyle{\frac{1 }{\left( k-1\right) !}}}}p_{i}{{\displaystyle {%
\frac{d^{k}x^{i} }{dt^{k}}}}}-kH_{0}\left( x^{i},{{\displaystyle {\frac{%
dx^{i} }{dt}}}},\ldots,{{\displaystyle{\frac{1 }{\left( k-1\right) !}}}}{{%
\displaystyle{\frac{d^{k-1}x^{i} }{dt^{k-1}}}}},p_{i}\right) =$ \\ 
${{\displaystyle{\frac{1 }{\left( k-1\right) !}}}}p_{i}{{\displaystyle {%
\frac{d^{k}x^{i} }{dt^{k}}}}}-k{{\displaystyle{\frac{\partial H_{0} }{%
\partial y^{(1)i}}}}}{{\displaystyle{\frac{dx^{i} }{dt}}}}-k{{\displaystyle {%
\frac{\partial H_{0} }{\partial y^{(2)i}}}}}{{\displaystyle{\frac{d^{2}x^{i} 
}{dt^{2}}}}}-\cdots$ \\ 
$-{{\displaystyle{\frac{k }{\left( k-1\right) !}}}}{{\displaystyle {\frac{%
\partial H_{0} }{\partial y^{(k-1)i}}}}}{{\displaystyle{\frac {d^{k-1}x^{i} 
}{dt^{k-1}}}}},$%
\end{tabular}
\end{equation*}
thus 
\begin{equation}
H_{0}=\overset{k-1}{\Gamma}{}(H_{0}),   \label{0eqzer1}
\end{equation}
where $\overset{k-1}{\Gamma}{}=y^{(1)i}{{\displaystyle{\frac{\partial }{%
\partial y^{(1)i}}}}}+\cdots+(k-1)y^{(k-1)i}{{\displaystyle{\frac{\partial }{%
\partial y^{(k-1)i}}}}}$ is a Liouville vector field. Notice that along the
curve $\gamma$ we have $y^{(\alpha)i}={{\displaystyle{\frac{1 }{\alpha!}}}}{{%
\displaystyle{\frac{d^{\alpha}x^{i} }{dt^{\alpha}}}}}$, $\alpha =\overline{%
1,k-1}$.

Differentiating the both sides of the equality (\ref{0eqzerm}) with respect
to ${{\displaystyle{\frac{d^{2}\tilde{t} }{dt^{2}}}}}$, then taking $t=%
\tilde{t}$, we have 
\begin{align*}
0 & =p_{i}{{{{{{{{{{{{\binom{k }{2}}}}}}}}}}}}}{{\displaystyle{\frac{1 }{%
(k-1)!}}}}{{\displaystyle{\frac{d^{k-1}x^{i} }{dt^{k-1}}}}}-k{{\displaystyle 
{\frac{\partial H_{0} }{\partial y^{(2)i}}}}}{{\displaystyle{\frac{1 }{2!}}}}%
{{{{{{{{{{{{\binom{2 }{2}}}}}}}}}}}}}{{\displaystyle{\frac{dx^{i} }{dt}}}}-k{%
{\displaystyle{\frac{\partial H_{0} }{\partial y^{(3)i}}}}}{{\displaystyle {%
\frac{1 }{3!}}}}{{{{{{{{{{{{\binom{3 }{2}}}}}}}}}}}}}{{\displaystyle {\frac{%
d^{2}x^{i} }{dt^{2}}}}}-\cdots \\
& -k{{\displaystyle{\frac{\partial H_{0} }{\partial y^{(k-2)i}}}}}{{%
\displaystyle{\frac{1 }{\left( k-1\right) !}}}}{{{{{{{{{{{{\binom{k-1 }{2}}}}%
}}}}}}}}}{{\displaystyle{\frac{d^{k-2}x^{i} }{dt^{k-2}}}}},
\end{align*}
thus 
\begin{equation*}
\left( k-1\right) p_{i}y^{(k-1)i}=\overset{k-2}{\Gamma}{}(H_{0}), 
\end{equation*}
where $\overset{k-2}{\Gamma}{}=y^{(1)i}{{\displaystyle{\frac{\partial }{%
\partial y^{(2)i}}}}}+\cdots+(k-1)y^{(k-2)i}{{\displaystyle{\frac{\partial }{%
\partial y^{(k-1)i}}}}}$ is a Liouville vector field.

Successively, differentiating the both sides of the equality (\ref{0eqzerm})
with respect to ${{\displaystyle{\frac{d^{3}\tilde{t} }{dt^{3}}}}}$,$\ldots$%
, ${{\displaystyle{\frac{d^{k-1}\tilde{t} }{dt^{k-1}}}}}$, then taking $t=%
\tilde{t}$, one obtain: 
\begin{equation}
\alpha p_{i}y^{(\alpha)i}=\overset{\alpha-1}{\Gamma}{}(H_{0}),(\forall
)\alpha=\overline{2,k-1}.   \label{0eqzer2}
\end{equation}

Let us assume that $k\geq2$. If we differentiate the both sides of the
equality (\ref{0eqzerm}) with respect to ${{\displaystyle{\frac{d^{k}\tilde {%
t}}{dt^{k}}}}}$, then taking $t=\tilde{t}$, we obtain the condition that the
curve $\gamma$ must fulfill: 
\begin{equation}
p_{i}{{\displaystyle{\frac{dx^{i}}{dt}}}}=0.   \label{0eqzer3}
\end{equation}
Thus the integral action is dependent of the parametrization of curves,
provided that the curve do not satisfy condition (\ref{0eqzer3}).

The equations (\ref{0eqzer1}) and (\ref{0eqzer2}) constitute the set of 
\emph{Zermelo conditions} for the affine hamiltonian $h$, while (\ref%
{0eqzer3}) is the Zermelo condition for the curve. It follows that, in
general, we have the following conclusion.

\begin{pr}
\label{prZer}The integral action (\ref{0eqig}) depends on a parametrization
of a curve, provided that the curve do not satisfy the condition (\ref%
{0eqzer3}), thus there are no Zermelo conditions for an affine hamiltonian.
\end{pr}

\section{Lagrangians of order $k$}

Let us consider a lagrangian $L$ of order $k$ on $M$, thus $L:\widetilde {%
T^{k}M}\rightarrow I\!\!R$ is differentiable. We say that $L$ is \emph{%
hyperregular} if its vertical hessian $\left( {{\displaystyle
{\frac{\partial^{2}L}{\partial y^{(k)i}\partial y^{(k)j}}}}}\right) $ is
non-degenerate and the Legendre map $\mathcal{L}:$ $\widetilde{T^{k}M}%
\rightarrow\widetilde{T^{k\ast}M}$, $(x^{i},y^{(1)i},\ldots,y^{(k)i})\overset%
{\mathcal{L}}{\rightarrow}{}(x^{i},y^{(1)i},\ldots,y^{(k-1)i},{{\displaystyle%
{\frac{\partial L}{\partial y^{(k)i}}}}})$ is a global diffeomorphism.

The Legendre map defines an $\mathcal{L}$-morphism of the vertical vector
bundles $V\widetilde{T^{k}M}\rightarrow V\widetilde{T^{k\ast}M}$ (called the 
\emph{vertical Legendre morphism}) and expressed in local coordinates on
fibers by $(y^{(k)i},Y^{j})\rightarrow({{{\displaystyle{\frac{\partial L}{%
\partial y^{(k)i}}}}}},Y^{j}{{{\displaystyle{\frac{\partial^{2}L}{\partial
y^{(k)j}y^{(k)k}}}}}})$. The following result is classical. It is stated in 
\cite[Chap.3. Sect. 1.4]{AG1} for the euclidean case $M=I\!\!R^{l}$, then
for a manifold $M$, only for convex lagrangians, for a local lagrangian.

\begin{theor}
\label{thlegendre}a) Let $L:T^{k}M\rightarrow I\!\!R$ be a hyperregular
lagrangian of order $k$ on $M$. Considering local coordinates, let $%
(x^{i},y^{(1)i},\ldots,y^{(k-1)i},p_{i})\overset{\mathcal{L}^{-1}}{%
\rightarrow}{}H^{j}(x^{i},$ $y^{(1)i},\ldots,$ $y^{(k-1)i},$ $p_{i})$ be the
local form of the inverse $\mathcal{L}^{-1}$ of the Legendre map. Then the
local functions given by 
\begin{align}
H_{0}(x^{i},y^{(1)i},\ldots,y^{(k-1)i},p_{i})=p_{j}H^{j}(x^{i},y^{(1)i},%
\ldots,y^{(k-1)i},p_{i})-  \notag \\
\hspace*{-40mm}L(x^{i},y^{(1)i},\ldots,y^{(k-1)i},H^{i}(x^{i},y^{(1)i},%
\ldots,y^{(k-1)i},p_{i}))  \notag
\end{align}
define a hyperregular affine hamiltonian of order $k$ on $M$ and the
vertical Legendre morphism is an isometry .

b) Conversely, let $h$ be a hyperregular affine hamiltonian of order $k$ on $%
M$. Considering local coordinates, let $(x^{i},y^{(1)i},\ldots,y^{(k)i})%
\overset{\mathcal{H}^{-1}}{\rightarrow}{}$ $(x^{i},y^{(1)i},\ldots
,y^{(k-1)i},$ $L_{i}(x^{i},y^{(1)i},\ldots,y^{(k)i}))$ be the local form of
the inverse $\mathcal{H}^{-1}$ of the Legendre$^{\ast}$ map. Then the local
functions given by 
\begin{align}
L(x^{i},y^{(1)i},\ldots,y^{(k)i})=y^{(k)j}L_{j}(x^{i},y^{(1)i},\ldots
,y^{(k)i})-  \notag \\
\hspace*{-40mm}H_{0}(x^{i},y^{(1)i},\ldots,y^{(k-1)i},L_{i}(x^{i},y^{(1)i},%
\ldots,y^{(k)i})  \notag
\end{align}
define a global hyperregular lagrangian of order $k$ on $M$ and the vertical
Legendre$^{\ast}$ morphism is an isometry .

c) The constructions a) and b) are inverse each to the other.
\end{theor}

We say that the hyperregular lagrangian and the affine hamiltonian that
corresponds according to Theorem \ref{thlegendre} are \emph{dual} each to
the other.

Let us apply this constructions to convex affine hamiltonians and lagrangians

We say that the affine hamiltonian $h$ of order $k$ on $M$ is positively
defined (or \emph{convex} according to \cite{AG1}) if the matrix $\left( 
\newline
{\displaystyle{\frac{-\partial^{2}H_{0} }{\partial p_{i}\partial p_{j}}}}%
\right) $ is positively defined. It follows that the Legendre map $\mathcal{H%
}^{\ast}$ is a local diffeomorphism, but we suppose in that follows that $%
\mathcal{H}^{\ast}$ is a global diffeomorphism, i.e. $h$ is hyperregular. We
denote by $L$ the dual lagrangian given by Theorem \ref{thlegendre}. It is
easy to see that $L$ is positively defined.

\begin{pr}
\label{prpoz}

\begin{enumerate}
\item Assume that the lagrangian $L$ is hyperregular and positively defined
and let $h$ be its dual hyperregular affine hamiltonian. Then $h$ is
positively defined and the local functions $H_{0}$ of $h$ are given by 
\begin{equation*}
H_{0}(x^{i},y^{(1)i},\ldots,,y^{(k-1)i},p_{i})=%
\mathrel{\mathop{\max}\limits_{(y^{(k)i})\in
I\!\!R^{n}}}{}\left( p_{i}y^{(k)i}-L(x^{i},y^{(1)i},\ldots,y^{(k)i})\right) 
\end{equation*}
and the maximum is taken for $y^{(k)i}=H^{i}(x^{i},y^{(1)i},\ldots
,,y^{(k-1)i},p_{i})$.

\item Assume that the hamiltonian $h$ is hyperregular and positively defined
and let $L$ be its dual hyperregular affine lagrangian. Then $L$ is
positively defined and if $H_{0}$ is a local function of $h$ then $L$ is
given locally by 
\begin{equation*}
L(x^{i},y^{(1)i},\ldots,,y^{(k)i})=%
\mathrel{\mathop{\max}\limits_{(p_{i})\in
I\!\!R^{n}}}\left(
p_{i}y^{(k)i}-H_{0}(x^{i},y^{(1)i},\ldots,,y^{(k-1)i},p_{i})\right) 
\end{equation*}
and the maximum is taken for $p_{i}=L_{i}(x^{i},y^{(1)i},\ldots,y^{(k)i})$.
\end{enumerate}
\end{pr}

\emph{Proof.} 1. The positivity of $h$ follows from the fact that the
hessians of $L$ and its dual $h$ are inverse one to the other. The real
function $I\!\!R^{n}\ni y^{(k)i}\overset{\Phi}{\rightarrow}%
{}p_{i}y^{(k)i}-L(x^{i},y^{(1)i},\ldots,,y^{(k)i})\in I\!\!R$ is concave and
according to Fermat principle it has a unique maximum given by the only root 
$H^{i}$ of ${\displaystyle{\frac{\partial\Phi}{\partial y^{(k)i}}}}=0 $. But 
${\displaystyle{\frac{\partial\Phi}{\partial y^{(k)i}}}}=$ $p_{i}-{%
\displaystyle{\frac{\partial L }{\partial y^{(k)i}}}}(x^{i},y^{(1)i},%
\ldots,y^{(k)i})$. It follows that the maximum is taken for $%
y^{(k)i}=H^{i}(x^{i},y^{(1)i},\ldots,y^{(k-1)i},p_{i})$.

2. It follows by duality. q.e.d.

Notice that, since $H_{0}$ is only local a function, the considerations
performed in \cite[Ch.3, Sect.1.4]{AG1} have only a local validity.

Let $H$ be a vectorial hamiltonian of order $k$ on $M$ and $%
s:T^{k-1}M\rightarrow T^{k}M$ be an affine section. Then $H$ and $s$ define
canonically an affine hamiltonian $h$ of order $k$ on $M$ (Proposition \ref%
{prcorham}). If $H$ is hyperregular, then $h$ is also hyperregular. In the
case when the Legendre$^{\ast}$ map of $h$ is a diffeomorphism, then $h$
defines a lagrangian $L$ of order $k$ on $M$. In this way, as in \cite%
{Mi-ha-B}, a non-canonical one to one correspondences between lagrangians
and vectorial hamiltonians follows, via Legendre maps; it is non-canonical
since it depends essentially on the section $s$.

For any curve $\gamma:[0,1]\rightarrow M$ , we define the \emph{integral
action} of $L$ along the curve $\gamma$ by the formula 
\begin{equation*}
I(\gamma)=\int_{0}^{1}L\left( x^{i},{{\displaystyle{\frac{dx^{i}}{dt}}}}%
,\ldots,{{\displaystyle{\frac{1}{k!}}}}{{\displaystyle{\frac{d^{k}x^{i}}{%
dt^{k}}}}}\right) dt, 
\end{equation*}
where $\gamma$ has the local form $t\overset{\gamma}{\rightarrow}{}(x^{i}(t))
$. The critical condition (or Fermat condition in the case of an extremum)
can be used for the integral action using similar arguments as in \cite[%
Chapter 5]{Mi-ha-B} (where it is used in the case of an extremun and it is
called an \emph{extremum condition}). It gives the Euler-Lagrange equation.
A curve $\gamma:[0,1]\rightarrow M$, $\gamma(t)=(x^{i}(t))$ is a solution
curve of the Euler-Lagrange equation of the Lagrangian $L$ if 
\begin{equation}
{{{\displaystyle{\frac{\partial L}{\partial x^{i}}}}}}-{{\displaystyle {%
\frac{1}{1!}}}}{{{\displaystyle{\frac{d}{dt}}}}{{\displaystyle{\frac{%
\partial L}{\partial y^{(1)i}}}}}}+\cdots+(-1)^{k}{{\displaystyle{\frac{1}{k!%
}}}}{{{\displaystyle{\frac{d^{k}}{dt^{k}}}}}}{{{\displaystyle{\frac{\partial
L}{\partial y^{(k)i}}}}}}=0,   \label{Eu-La}
\end{equation}
along the curve $\gamma^{(k)}:I\rightarrow T^{k}M$, $%
\gamma^{(k)}(t)=(x^{i}(t),{{\displaystyle{\frac{1}{1!}}}}{{{\displaystyle{%
\frac{dx^{i}}{dt}}}}}(t),\ldots,{{\displaystyle{\frac{1}{k!}}}}{{{%
\displaystyle
{\frac{d^{k}x^{i}}{dt^{k}}}}}}(t))$. This equation is expressed in terms of
the Jacobi-Ostrogradski momenta in \cite{Mi-ha-B,AG1}. Here these
Jacobi-Ostrogradski momenta are $(p_{(0)i},\ldots,p_{(k-1)i})$, viewed here
as coordinates on $T^{\ast}\widetilde{T^{k}M}$.

Let us consider that the lagrangian $L$ is hyperregular and let $(x^{i}$, $%
y^{(1)i}$,$\ldots$, $y^{(k-1)i}$, $p_{i})\overset{\mathcal{L}^{-1}}{%
\rightarrow}{}(x^{i}$, $y^{(1)i}$,$\ldots$, $y^{(k-1)i}$, $%
H^{i}(x^{i},y^{(1)i},\ldots$, $y^{(k-1)i}$, $p_{i}))$ be the inverse of the
Legendre map. Thus ${{{\displaystyle{\frac{\partial L}{\partial y^{(k)i}}}}}}%
(x^{i}$, $y^{(1)i}$,$\ldots$, $y^{(k-1)i}$, $H^{i}(x^{i},y^{(1)i},%
\ldots,y^{(k-1)i}$, $p_{i}))=p_{i}$ and $H^{i}(x^{i},y^{(1)i},\ldots$, $%
y^{(k-1)i}$, ${{{\displaystyle{\frac{\partial L}{\partial y^{(k)i}}}}}}%
(x^{i},y^{(1)i},\ldots,y^{(k-1)i},y^{(k)i}))=y^{(k)i}$. It follows that ${{{%
\displaystyle{\frac{\partial H^{i}}{\partial p_{j}}}}}}=g^{ij}$. Let us
consider the affine hamiltonian $h$ corresponding to $L$, according to
Theorem \ref{thlegendre}. We call the \emph{total energy} of the
hyperregular lagrangian $L$ to be the energy function $\mathcal{E}:%
\widetilde{T^{\ast}T^{k}M}\rightarrow I\!\!R$ of $h$. Taking into account
the local form of $\mathcal{E}$ given by (\ref{eqh0}) and the definition of $%
H_{0}$ given by Theorem \ref{thlegendre}, we obtain: 
\begin{align}
\mathcal{E} & =p_{(0)i}y^{(1)i}+\cdots+(k-1)p_{(k-2)i}y^{(k-1)i}+
\label{eqh1} \\
& kp_{(k-1)i}H^{i}(x^{i},y^{(1)i},\ldots,y^{(k-1)i},p_{(k-1)i})-  \notag \\
& kL(x^{i},y^{(1)i},\ldots,y^{(k-1)i},H^{i}(x^{i},y^{(1)i},\ldots
,y^{(k-1)i},p_{(k-1)i})).  \notag
\end{align}

An integral curve of the hamiltonian vector field $X_{\mathcal{E}}$ is a
solution of the first order differential equations, determined by the
lagrangian $L$: 
\begin{equation*}
\left\{ 
\begin{array}{l}
{{{\displaystyle{\frac{dx^{i}}{dt}}}}}={{{\displaystyle{\frac{\partial 
\mathcal{E}}{\partial p_{(0)i}}}}}}, \\ 
{{{\displaystyle{\frac{dy^{(\alpha)i}}{dt}}}}}={{{\displaystyle{\frac {%
\partial\mathcal{E}}{\partial p_{(\alpha)i}}}}}},\alpha=\overline{1,k-1}, \\ 
{{{\displaystyle{\frac{dp_{(\alpha)i}}{dt}}}}}=-{{{\displaystyle {\frac{%
\partial\mathcal{E}}{\partial y^{(\alpha)i}}}}}},\alpha=\overline {0,k-1}.%
\end{array}
\right. 
\end{equation*}
These are called as \emph{Hamilton-Jacobi equations} in \cite{Mi-Lk} and as 
\emph{generalized Euler-Lagrange dynamical equations }in \cite{Le}. Notice
that $\mathcal{E}$ has different meanings in \cite{Mi-Lk, Le}.

Thus an integral curve of the vector field $X_{\mathcal{E}}$ is a solution
of the differential equations: 
\begin{equation}
\left\{ 
\begin{array}{l}
{{{\displaystyle{\frac{dx^{i}}{dt}}}}}=y^{(1)i}, \\ 
{{{\displaystyle{\frac{dy^{(\alpha)i}}{dt}}}}}=(\alpha+1)y^{(\alpha
+1)i},\alpha=\overline{1,k-2}, \\ 
{{{\displaystyle{\frac{dy^{(k-1)i}}{dt}}}}}=kH^{i}, \\ 
{{{\displaystyle{\frac{dp_{(0)i}}{dt}}}}}=k{{{\displaystyle{\frac{\partial L%
}{\partial x^{i}}}}}}, \\ 
{{{\displaystyle{\frac{dp_{(\alpha)i}}{dt}}}}}=k{{{\displaystyle {\frac{%
\partial L}{\partial y^{(\alpha)i}}}}}}-\alpha p_{(\alpha-1)i},\alpha=%
\overline{1,k-1},%
\end{array}
\right.   \label{eqjhnew}
\end{equation}
if $k\geq2$ and 
\begin{equation*}
\left\{ 
\begin{array}{l}
{{{\displaystyle{\frac{dx^{i}}{dt}}}}}=H^{i}(x^{j},p_{(0)j}), \\ 
{{{\displaystyle{\frac{dp_{(0)i}}{dt}}}}}={{{\displaystyle{\frac{\partial L}{%
\partial x^{i}}}}}}\left( x^{i},H^{i}(x^{j},p_{(0)j})\right) ,%
\end{array}
\right. 
\end{equation*}
in the case $k=1$. This case recovers the Hamilton equation: 
\begin{equation*}
\left\{ 
\begin{array}{l}
{{{\displaystyle{\frac{dx^{i}}{dt}}}}}={{{\displaystyle{\frac{\partial 
\mathcal{E}}{\partial p_{i}}}}}}, \\ 
{{{\displaystyle{\frac{dp_{(0)i}}{dt}}}}}=-{{{\displaystyle{\frac {\partial%
\mathcal{E}}{\partial x^{i}}}}}},%
\end{array}
\right. 
\end{equation*}
where $\mathcal{E}%
(x^{j},p_{(0)j})=p_{i}H^{i}(x^{j},p_{(0)j})-L(x^{j},H^{i}(x^{j},p_{(0)j}))$
is the energy of $L$ and $\mathcal{E}$ is viewed as a hamiltonian of order $%
1 $ on $M$.

Similar to the hamiltonian case (Theorem \ref{thgeoham}), one has the
following result.

\begin{theor}
\label{thgeolagr}Assume that the lagrangian $L$ is hyperregular. Then the
integral curves of the hamiltonian vector field $X_{\mathcal{E}}$ projects
on curves on $M$ which are solutions of the Euler-Lagrange equation of $L$.
\end{theor}

\emph{Proof.} Let us consider the fibered product $\mathcal{T}^{k}M=T^{\ast
}T^{k-1}M\times_{T^{k-1}M}T^{k}M$, where both the fibered manifolds $T^{\ast
}T^{k-1}M\rightarrow T^{k-1}M$ (a vector bundle) and $T^{k}M\rightarrow
T^{k-1}M$ (an affine bundle) are considered over the same base, $T^{k-1}M$.
Consider some local coordinates: $(x^{i}$, $y^{(1)i}$,$\ldots$, $y^{(k-1)i})$
on $T^{k-1}M$, $(x^{i}$, $y^{(1)i}$,$\ldots$, $y^{(k-1)i}$, $y^{(k)i})$ on $%
T^{k}M$, $(x^{i}$, $y^{(1)i}$,$\ldots$, $y^{(k-1)i}$, $p_{(0)i},%
\ldots,p_{(k-1)i})$ on $T^{\ast}T^{k-1}M$ and $(x^{i}$, $y^{(1)i}$,$\ldots$, 
$y^{(k-1)i}$, $y^{(k)i}$, $p_{(0)i},\ldots,p_{(k-1)i})$ on $\mathcal{T}^{k}M$%
. The hyperregular lagrangian $L$ defines a canonical embedding $I:$ $%
T^{\ast }T^{k-1}M\rightarrow\mathcal{T}^{k}M$ given in local coordinates by $%
(x^{i}$, $y^{(1)i}$,$\ldots$, $y^{(k-1)i}$, $p_{(0)i},\ldots,p_{(k-1)i})%
\overset{I}{\rightarrow}{}(x^{i}$, $y^{(1)i}$,$\ldots$, $y^{(k-1)i}$, $%
H^{i}(x^{i},y^{(1)i},\ldots,y^{(k-1)i},p_{(k-1)i})$, $p_{(0)i},%
\ldots,p_{(k-1)i})$, where $H^{i}$ is the local form of the inverse $%
\mathcal{L}^{-1}$ of the Legendre map $\mathcal{L}$ of $L$, considered
previously. The submanifold $I(T^{\ast}T^{k-1}M)\subset\mathcal{T}^{k}M$ is
given also by the local equation $p_{(k-1)i}={{{\displaystyle{\frac{\partial
L}{\partial y^{(k)i}}}}}}(x^{i}$, $y^{(1)i}$,$\ldots$, $y^{(k)i})$ on $%
\mathcal{T}^{k}M$. Using the system (\ref{eqjhnew}) one obtain that the
equation (\ref{Eu-La}) is satisfied along the integral curves of the
hamiltonian vector field $X_{\mathcal{E}}$. q.e.d.

A lagrangian and its dual hamiltonian are related as follows.

\begin{theor}
\label{the}Assume that the lagrangian $L$ or its dual affine hamiltonian $h$
is hyperregular. Then $L$ and $h$ have the same energy thus they have the
same hamiltonian vector field on $T^{k-1}M$.
\end{theor}

\emph{Proof.} The fact that $L$ and $h$ have the same energy follows from
formulas (\ref{eqh0}) and (\ref{eqh1}), using the form of Legendre map given
by Theorem \ref{thlegendre}. q.e.d.

\begin{cor}
\label{corlagham}Assume that the lagrangian $L$ or its dual affine
hamiltonian $h$ is hyperregular. Then the curves on $M$ that are solutions
of the Euler-Lagrange equation of $L$ are the same as the projections on $M$
of the curves on $T^{\ast}M$ that are solutions of the Hamilton equation of $%
h$.
\end{cor}

We consider now some non-trivial examples. For sake of simplicity we take $%
k=2$.

Let $L:TM\rightarrow I\!\!R$ be a lagrangian. Let us consider $%
L^{(2)}:T^{2}M\rightarrow I\!\!R$, $L^{(2)}(x^{i},y^{(1)i},y^{(2)i})=$ $%
\varepsilon _{0}L(x^{i},y^{(1)i})+\varepsilon_{1}L(x^{i},z^{(2)i})$, where $%
\varepsilon _{0}$, $\varepsilon_{1}\in I\!\!R$, $\varepsilon_{1}\neq0$, $%
z^{(2)i}=y^{(2)i}-S^{i}(x^{i},y^{(1)i})$ and $2S^{i}=g^{il}\left( y^{l}{%
\displaystyle{\frac{\partial^{2}L }{\partial x^{l}y^{j}}}}-{\displaystyle
{\frac{\partial L }{\partial x^{j}}}}\right) $ are the components of the
semi-spray of $L$ . We denote by $H:T^{\ast}M\rightarrow I\!\!R$ the dual
hamiltonian that correspunds to $L$, i.e. $%
H(x^{i},p_{i})=p_{i}H^{i}(x^{i},p_{i})-L(x^{i},H^{i})$, where ${\displaystyle%
{\frac{\partial L }{\partial y^{j}}}}(x^{i},H^{i}(x^{i},p_{i}))={}p_{j}$.

We have ${\displaystyle{\frac{\partial L^{(2)} }{\partial y^{(2)i}}}}%
=\varepsilon_{1}{\displaystyle{\frac{\partial L }{\partial y^{i}}}}%
(x^{j},y^{(2)j}-S^{j}(x^{j},y^{(1)j}))=p_{i}$, thus \newline
$y^{(2)i}=S^{i}(x^{j},y^{(1)j})+H^{i}(x^{j},{\displaystyle{\frac{1 }{%
\varepsilon_{1}}}}p_{j})\overset{not.}{=}{}h^{i}(x^{j},y^{(1)j},p_{j})$. It
follows that the dual affine hamiltonian that corresponds to $L^{(2)}$ is 
\newline
$%
2H^{(2)}(x^{j},y^{(1)j},p_{j})=p_{i}h^{i}(x^{j},y^{(1)j},p_{j})-L^{(2)}(x^{j},y^{(1)j},h^{j})=p_{i}S^{i}(x^{j},y^{(1)j})+ 
$ $p_{i}H^{i}(x^{j},{\displaystyle{\frac{1 }{\varepsilon_{1}}}}p_{j})-$ $%
\varepsilon _{0}L(x^{i},y^{(1)i})-$ $\varepsilon_{1}L(x^{i},H^{i}(x^{j},{%
\displaystyle
{\frac{1 }{\varepsilon_{1}}}}p_{j}))=$ \newline
$p_{i}S^{i}(x^{j},y^{(1)j})+$ $\varepsilon_{1}H(x^{j},{\displaystyle{\frac{1 
}{\varepsilon_{1}}}}p_{j})-\varepsilon_{0}L(x^{j},y^{(1)j})$.

The energy of $L^{(2)}$ is $\mathcal{E}%
^{(2)}(x^{i},y^{(1)i},p_{(0)i},p_{(1)i})=p_{(0)i}y^{(1)i}+2H^{(2)}(x^{i},y^{(1)i}, 
$ $p_{(1)i}) $, or $\mathcal{E}%
^{(2)}(x^{i},y^{(1)i},p_{(0)i},p_{(1)i})=p_{(0)i}y^{(1)i}+p_{(1)i}S^{i}(x^{j},y^{(1)j})+ 
$\newline
$\varepsilon_{1}H(x^{j},{\displaystyle{\frac{1 }{\varepsilon_{1}}}}p_{(1)j})-
$ $\varepsilon _{0}L(x^{j},y^{(1)j}). $

The hamiltonian field of $\mathcal{E}^{(2)}$ is $X_{\mathcal{E}^{(2)}}={%
\displaystyle{\frac{\partial\mathcal{E}^{(2)} }{\partial p_{(0)i}}}}{%
\displaystyle{\frac{\partial}{\partial x^{i}}}}+{\displaystyle
{\frac{\partial\mathcal{E}^{(2)} }{\partial p_{(1)i}}}}{\displaystyle
{\frac{\partial}{\partial y^{(1)i}}}}-$ \newline
${\displaystyle
{\frac{\partial\mathcal{E}^{(2)} }{\partial x^{i}}}}{\displaystyle
{\frac{\partial}{\partial p_{(0)i}}}}-{\displaystyle{\frac{\partial \mathcal{%
E}^{(2)} }{\partial y^{(1)i}}}}{\displaystyle{\frac{\partial }{\partial
p_{(1)i}}}}$. Thus the integral curves of the hamiltonian vector field $X_{%
\mathcal{E}^{(2)}}$ are given by the equations: 
\begin{equation*}
\left\{ 
\begin{array}{l}
{\displaystyle{\frac{dx^{i} }{dt}}} =y^{(1)i}, \\ 
{\displaystyle{\frac{dy^{(1)i} }{dt}}} =S^{i}(x^{j},y^{(1)j})+{\displaystyle 
{\frac{\partial H }{\partial p_{i}}}}(x^{j},{\displaystyle{\frac{1 }{%
\varepsilon_{1}}}}p_{(1)j}), \\ 
{\displaystyle{\frac{dp_{(0)i} }{dt}}} =-p_{(1)j}{\displaystyle{\frac{%
\partial S^{j} }{\partial x^{i}}}}(x^{j},y^{(1)j})-\varepsilon_{1}{%
\displaystyle {\frac{\partial H }{\partial x^{i}}}}(x^{j},{\displaystyle{%
\frac{1 }{\varepsilon_{1}}}}p_{(1)j})+\varepsilon_{0}{\displaystyle{\frac{%
\partial L }{\partial x^{i}}}}(x^{j},y^{(1)j}), \\ 
{\displaystyle{\frac{dp_{(1)i} }{dt}}} =-p_{(0)i}-p_{(1)j}{\displaystyle {%
\frac{\partial S^{j} }{\partial y^{(1)i}}}}(x^{j},y^{(1)j})+\varepsilon _{0}{%
\displaystyle{\frac{\partial L }{\partial y^{(1)i}}}}(x^{j},y^{(1)j}).%
\end{array}
\right. 
\end{equation*}

If we denote ${\displaystyle{\frac{1 }{\varepsilon_{1}}}}p_{(1)i}=z_{i}$, we
have: ${\displaystyle{\frac{1 }{\varepsilon_{1}}}}p_{(0)i}={\displaystyle
{\frac{\varepsilon_{0} }{\varepsilon_{1}}}}{\displaystyle{\frac{\partial L }{%
\partial y^{(1)i}}}}(x^{j},y^{(1)j})- $\newline
$z_{j}{\displaystyle
{\frac{\partial S^{j} }{\partial y^{j}}}}(x^{j},y^{(1)j})-{\displaystyle
{\frac{dz_{i} }{dt}}}$, thus \newline
${\displaystyle{\frac{\varepsilon_{0} }{\varepsilon_{1}}}}(y^{(1)j}{%
\displaystyle{\frac{\partial^{2}L }{\partial x^{j}\partial y^{i}}}}%
(x^{j},y^{(1)j})+(H^{j}(x^{j},z_{j})+S^{j}(x^{j},y^{(1)j})){\displaystyle{%
\frac{\partial^{2}L }{\partial y^{j}\partial y^{i}}}}(x^{j},y^{(1)j}))$ $-{%
\displaystyle{\frac{d^{2}z_{i} }{dt^{2}}}}=$ ${\displaystyle{\frac{%
\varepsilon_{0} }{\varepsilon_{1}}}}{\displaystyle
{\frac{\partial L }{\partial x^{i}}}}(x^{j},y^{(1)j})-z_{j}{\displaystyle
{\frac{\partial S^{j} }{\partial x^{i}}}}(x^{j},y^{(1)j})-{\displaystyle
{\frac{\partial H }{\partial x^{i}}}}(x^{j},z_{j})$ and ${{\displaystyle
{\frac{dy^{(1)i} }{dt}}}}=H^{i}(x^{j},z_{i})+S^{i}(x^{j},y^{(1)j})$, thus $%
z_{i}={\displaystyle{\frac{\partial L }{\partial x^{i}}}}(x^{j},{{%
\displaystyle{\frac{dy^{(1)j} }{dt}}}-}S^{j}(x^{k},y^{(1)k}))$.

Discussion:

If ${\displaystyle{\frac{\varepsilon_{0} }{\varepsilon_{1}}}}\rightarrow0 $
(or even when $\varepsilon_{0}=0$), it follows the equation \newline
${\displaystyle{\frac{d^{2}z_{i} }{dt^{2}}}}=z_{j}{\displaystyle
{\frac{\partial S^{j} }{\partial x^{i}}}}(x^{j},y^{(1)j})+{\displaystyle
{\frac{\partial H }{\partial x^{i}}}}(x^{j},z_{j})$.

If ${\displaystyle{\frac{\varepsilon_{1} }{\varepsilon_{0}}}}\rightarrow0 $,
it follows the equation \newline
$y^{(1)j}{\displaystyle{\frac{\partial^{2}L }{\partial x^{j}\partial y^{i}}}}%
(x^{j},y^{(1)j})+(H^{j}(x^{j},z_{j})+S^{j}(x^{j},y^{(1)j})){\displaystyle{%
\frac{\partial^{2}L }{\partial y^{j}\partial y^{i}}}}(x^{j},y^{(1)j})= $%
\newline
${\displaystyle
{\frac{\partial L }{\partial x^{i}}}}(x^{j},y^{(1)j})$, thus $%
H^{j}(x^{j},z_{j})=0$.

Let us suppose that the equation $H^{i}(x^{j},z_{j})=0$ has as solution $%
z_{j}=\varphi_{j}(x^{i})$ and $(x^{j},y^{j})\rightarrow{\displaystyle
{\frac{\partial L }{\partial y^{i}}}}(x^{j},y{^{j})}$ is continuous in $%
(x^{j},y^{j}=0)$. Then since ${\displaystyle{\frac{\partial L }{\partial
y^{i}}}}(x^{j},H^{i}(x^{j},z_{j}){)=z}_{j}$, it follows that $\varphi
_{j}(x^{i})={\displaystyle{\frac{\partial L }{\partial y^{j}}}}%
(x^{i},H^{i}(x^{i},\varphi_{i}(x^{i})){)=}$ \newline
${\displaystyle{\frac{\partial L }{\partial y^{j}}}}(x^{i},0)$. Thus along a
geodesic of $L$, taking $z_{j}=\varphi_{j}(x^{i})=\xi_{(1)j}(x^{i},0)$, we
have $H^{j}(x^{j},z_{j})=0$.

Let us consider some particular cases.

\textbf{Example 1.} In the canonical euclidian plane $E^{2}$ we consider the
lagrangian $L(x^{i},y^{i})=L(x^{1},x^{2},y^{1},y^{2})={\displaystyle{\frac {1%
}{2}}}((y^{1})^{2}+(y^{2})^{2})$. We have $S^{1}=S^{2}=0$ and the dual
hamiltonian of $L$ is $H(x^{i},p_{i})={\displaystyle{\frac{1}{2}}}%
((p_{1})^{2}+(p_{2})^{2})$. The second order energy is $\mathcal{E}%
^{(2)}(x^{i},y^{(1)i},p_{(0)i},p_{(1)i})=$ $p_{(0)i}y^{(1)i}+$ ${%
\displaystyle
{\frac{1}{2\varepsilon_{1}}}}((p_{(1)1})^{2}+(p_{(1)2})^{2})$ $-{%
\displaystyle
{\frac{\varepsilon_{0}}{2}}}((y^{(1)1})^{2}+(y^{(1)2})^{2}).$

The Hamilton equation is: 
\begin{equation*}
\left\{ 
\begin{array}{l}
{\displaystyle{\frac{dx^{i} }{dt}}}={\displaystyle{\frac{\partial \mathcal{E}%
^{(2)} }{\partial p_{(0)i}}}}=y^{(1)i}, \\ 
{\displaystyle{\frac{dy^{(1)i} }{dt}}}={\displaystyle{\frac{\partial 
\mathcal{E}^{(2)} }{\partial p_{(1)i}}}}={\displaystyle{\frac{1 }{%
\varepsilon_{1}}}}p_{(1)i}, \\ 
{\displaystyle{\frac{dp_{(0)i} }{dt}}}=-{\displaystyle{\frac{\partial 
\mathcal{E}^{(2)} }{\partial x^{i}}}}=0, \\ 
{\displaystyle{\frac{dp_{(1)i} }{dt}}}=-{\displaystyle{\frac{\partial 
\mathcal{E}^{(2)} }{\partial y^{(1)i}}}}=-p_{(0)i}+\varepsilon_{0}y^{(1)i}.%
\end{array}
\right. 
\end{equation*}

It follows that ${\displaystyle{\frac{d^{4}x^{i} }{dt^{4}}}}+\alpha {%
\displaystyle{\frac{d^{2}x^{i} }{dt^{2}}}}=0$, where $\alpha=-{%
\displaystyle
{\frac{\varepsilon_{0} }{\varepsilon_{1}}}}$. The solutions of this
differential equation can be easy found. We analyse briefly the general
solutions in different cases.

If $\alpha=0$, i.e. $\varepsilon_{0}=0$, the general solution has the form $%
x^{i}=a_{i}t^{3}+b_{i}t^{2}+c_{i}t+d_{i}$, where $a_{i}$, $b_{i}$, $c_{i}$
and $d_{i}\in I\!\!R$.

If $\alpha<0$, i.e. $\varepsilon_{0}\cdot\varepsilon_{1}>0$, the general
solution has the form $x^{i}=a_{i}\cos t\sqrt{\alpha}+b_{i}\sin t\sqrt{%
\alpha }+c_{i}t+d_{i}$, where $a_{i}$, $b_{i}$, $c_{i}$ and $d_{i}\in I\!\!R$%
.

If $\alpha>0$, i.e. $\varepsilon_{0}\cdot\varepsilon_{1}<0$, the general
solution has the form $x^{i}=a_{i}\cosh t\sqrt{-\alpha}+b_{i}\sinh t\sqrt{%
-\alpha}+c_{i}t+d_{i}$, where $a_{i}$, $b_{i}$, $c_{i}$ and $d_{i}\in I\!\!R 
$.

Notice that if $\varepsilon_{1}=\pm\varepsilon_{2}=\varepsilon\neq0$, the
general solution is the same; thus the ,,size'', given by $\varepsilon$ is
not important for the solution of the Hamilton equation.

\textbf{Example 2.} A similar discussion can be performed in the canonical
euclidian plane $E^{2}$, considering the lagrangian $%
L(x^{i},y^{i})=L((x^{1},x^{2}),(y^{1},y^{2}))={\displaystyle{\frac{1}{2}}}%
((y^{1})^{2}+(y^{2})^{2})+a_{1}y^{1}+a_{2}y^{2}$, where $a_{1}$, $a_{2}\in
I\!\!R$ are constants.

\textbf{Example 3.} In the canonical euclidian plane $E^{2}$ we consider a
function $V:E^{2}\rightarrow I\!\!R$ and the lagrangian $%
L(x^{i},y^{i})=L((x^{1},x^{2}),(y^{1},y^{2}))={\displaystyle{\frac{1}{2}}}%
((y^{1})^{2}+(y^{2})^{2})+V(x^{i})$. We have $H={\displaystyle{\frac{1}{2}}}%
((p_{1})^{2}+(p_{2})^{2})-V(x^{i})$ and $S^{i}=-{\displaystyle{\frac{1}{2}}}{%
\displaystyle{\frac{\partial V}{\partial x^{i}}}}$. The Hamilton equation
becomes: 
\begin{equation*}
\left\{ 
\begin{array}{l}
{\displaystyle{\frac{d^{2}x^{i}}{dt^{2}}}}=-{\displaystyle{\frac{1}{2}}}%
\sum\limits_{j}p_{j}{\displaystyle{\frac{\partial^{2}V}{\partial
x^{i}\partial x^{j}}}}-\left( \varepsilon_{0}+\varepsilon_{1}\right) {%
\displaystyle {\frac{\partial V}{\partial x^{i}}}}; \\ 
{\displaystyle{\frac{d^{2}p_{i}}{dt^{2}}}}=-{\displaystyle{\frac{1}{2}}}{%
\displaystyle{\frac{\partial V}{\partial x^{i}}}}+\varepsilon_{1}p_{i}.%
\end{array}
\right. 
\end{equation*}

Let us consider some particular cases for $V$.

The first case is $V(x^{i})=\alpha_{i}x^{i}$, where $\alpha_{i}$ are
constants and $\alpha_{1}^{2}+\alpha_{2}^{2}\neq0$. The first equation gives 
${\displaystyle{\frac{d^{2}x^{i} }{dt^{2}}}}=-{\displaystyle{\frac{1 }{2}}}%
\alpha_{i}\left( \varepsilon_{0}+\varepsilon_{1}\right) $, thus the solution
curves of Hamilton equation (critical curves) are parabolas if $%
\varepsilon_{0}+\varepsilon_{1}\neq0$ and stright lines if $\varepsilon
_{0}+\varepsilon_{1}=0$. Notice that the critical curves depend only on $%
\varepsilon_{0}+\varepsilon_{1}$.

The second case is $V(x^{i})=\sum\limits_{i}\left( x^{i}\right) ^{2}$, i.e. $%
V$ has a spherical symmetry. In this case the equation of critical curves
is: 
\begin{equation*}
\left\{ 
\begin{array}{l}
{\displaystyle{\frac{d^{2}x^{i}}{dt^{2}}}}=-2\left( \varepsilon
_{0}+\varepsilon _{1}\right) x^{i}-p_{i}; \\ 
{\displaystyle{\frac{d^{2}p_{i}}{dt^{2}}}}=-x^{i}+\varepsilon _{1}p_{i}.%
\end{array}%
\right. 
\end{equation*}%
The equation can be integrated as follows. Take $z_{i}=x^{i}+\beta p_{i}$
such that ${\displaystyle{\frac{d^{2}z_{i}}{dt^{2}}}}=\gamma z_{i}$, thus $%
-2(\varepsilon _{0}+\varepsilon _{1})-\beta =\gamma $ and $-1+\beta
\varepsilon _{1}=\gamma \beta $. It follows that $\beta ^{2}+\beta
(2\varepsilon _{0}+3\varepsilon _{1})-1=0$ with non-null real roots $\beta
_{1}\neq \beta _{2}$. We have $\gamma _{1,2}=-2(\varepsilon _{0}+\varepsilon
_{1})-\beta _{1,2}={\displaystyle{\frac{-(4\varepsilon _{0}+5\varepsilon
_{1})\pm \sqrt{(2\varepsilon _{0}+3\varepsilon _{1})^{2}+4}}{2}}}$, thus $%
\gamma _{1}\neq \gamma _{2}$. According to $\gamma _{1}$ and $\gamma _{2}$
one obtains the general solution that gives the equation of critical curves.
Let $f_{i}=x^{i}+\beta _{1}p_{i}$ and $g_{i}=x^{i}+\beta _{2}p_{i}$ be the
general solutions that corresponds to $\gamma _{1}$ and $\gamma _{2}$
repectively. We have that $x^{i}={\displaystyle{\frac{\beta _{2}f_{i}-\beta
_{1}g_{i}}{\beta _{2}-\beta _{1}}}}$ is the general solution of the equation
of critical curves in this case.

\textbf{Acknowledgements}

The authors were partially supported by the CNCSIS grant A No. 81/2005.

\noindent Authors' address:

{\large Paul Popescu} and {\large Marcela Popescu}\newline
\emph{University of Craiova}\newline
\emph{Department of Applied Mathematics}\newline
\emph{13, Al.I.Cuza st., Craiova, 200585, Romania}\newline
\emph{E-mail addresses: paul\_p\_popescu@yahoo.com, marcelacpopescu@yahoo.com%
}

\end{document}